\documentclass{aa}
\usepackage{graphics}
\def\hi{H\,{\sc i} }

\def\kmss{km~s$^{-1}$ }
\def\kms{km~s$^{-1}$}

\def\dispr{{\langle}v^2_R{\rangle}^{1/2}}
\def\disprn{{\langle}v^2_R{\rangle}^{1/2}_{R=0}}
\def\disprh{{\langle}v^2_R{\rangle}^{1/2}_{R=h}}
\def\dispt{{\langle}v^2_{\Theta}{\rangle}^{1/2}}
\def\dispz{{\langle}v^2_z{\rangle}^{1/2}}
\def\dispzn{{\langle}v^2_z{\rangle}^{1/2}_{R=0}}
\begin{document}
\thesaurus{03(11.09.1 NGC 7331;  09.11.1 11.06.2 11.11.1 11.19.2)} 
\title{The kinematics of the bulge and the disc of NGC 7331}
\author{Roelof Bottema}
\institute{Kapteyn Astronomical Institute, P.O. Box 800, 
NL-9700 AV Groningen, The Netherlands, e-mail:robot@astro.rug.nl}
\date{Received date1; accepted date2}
\maketitle
\begin{abstract}
Results are presented of spectroscopic emission and absorption line
observations along the major axis of the Sb galaxy NGC 7331. The
kinematics of the ionized gas and stellar component are derived, being
regular and symmetric with respect to the centre of the galaxy. 
Contrary to what may be expected, for $R$ $\la$ 40\arcsec\ the emission line
gas appears to rotate slower than the stars. A likely explanation
for this phenomenon is an inclined and warped gas layer
in those inner regions. In the bulge dominated region the absorption
line profiles are asymmetric in the sense that they have a shallow
extension towards the systemic velocity. No counterrotating stellar
component is observed which is contrary to previous
claims.
As demonstrated, these claims might be based on a wrong interpretation
of the employed analysis method.
Outside the bulge dominated region
the stellar radial velocities are in agreement with the neutral
hydrogen dynamics and the stellar velocity dispersion decreases
towards larger radii.

A detailed bulge/disc light decomposition has been made.
That has been used to construct a kinematical model
of NGC 7331 from which model absorption line
profiles were calculated. These profiles have been compared
with the observations and model parameters have been adjusted to
obtain a good match. It appeared necessary to
combine a rapidly rotating disc having a radially
decreasing velocity dispersion with a slowly rotating
constant dispersion bulge.
Then, simultaneously, the observed stellar radial velocities, the
velocity dispersions and the observed asymmetry of the line
profile could be explained satisfactorily. An even better fit
to the data can be achieved when the disc is relatively
thinner and colder inside the bulge region. 

For the disc the stellar velocity dispersions and photometry
result in a mass-to-light ratio of $1.6 \pm 0.7$ $M_{\odot}/L_{\odot}^I$.
This value agrees with previous determinations for other discs
using observed velocity dispersions.
A rotation curve analysis allows the calculation of the
mass-to-light ratio of the bulge which amounts to 6.8 in the I-band;
considerably larger than the disc value. It appears that the mass
distribution of NGC 7331 is completely dominated by the 
combination of bulge
and dark halo at all radii. Comparing well determined mass-to-light
ratios of a number of bulges with disc values, on average,
$(M/L)_{\rm bulge}$ is three times as large as
$(M/L)_{\rm disc}$ in the I-band. For the B-band this ratio goes
up to 7.2, a fact which should have cosmological consequences.
\end{abstract}
\section{Introduction}

Beyond the optical edge of disc galaxies large amounts of dark matter are 
needed to explain the flat rotation curves determined from
the velocity field of the \hi gas (Bosma \cite{bosma78}; Begeman
\cite{begeman87}). However,
the contribution of the luminous disc matter to the total
rotation cannot be determined from a rotation curve analysis
(van Albada et al. \cite{albada85}). On the other hand, knowing the surface
density of the disc allows the construction of the rotation curve
of the disc only. Comparison with the observed total rotation then
gives the rotational contribution of the disc. Galactic discs can
essentially be described as locally isothermal stellar sheets embedded
in a dark halo. Measuring the stellar velocity dispersion
of the disc then provides the surface density to determine
this rotational contribution. 

For a dozen galactic discs these velocity dispersion measurements
have been done (Bottema \cite{bottema93}, and references therein)
though for most of these only over a limited radial extent.
The general conclusion is that a galactic disc provides, at the 
position of the maximum rotation of the disc, 63\% $\pm$ 10\% of
the total rotation of a galaxy (Bottema \cite{bottema93}, hereafter B93).
Using an analysis of the light and colour distributions of galactic
discs this conclusion can be rewritten to $(M/L)_B = 1.8 \pm 0.4$
when $B-V = 0.7$. For a disc with a
certain brightness and colour a general expression is determined for
its rotational contribution and for its mass-to-light ratio
(Bottema \cite{bottema97a}). Unfortunately the error
on the mass-to-light ratio is still large caused by the small sample
of discs for which dispersions have actually been measured. 
Consequently an extension of the sample is badly needed.

There are other problems which can be addressed when disc dispersions
are better known, for instance galaxy formation. Nowadays it is believed
that a collapse of dark matter creates the potential well in which
baryons collect to form a galaxy (e.g. Katz \& Gunn \cite{katz91}).
To check if this scenario is right and to understand the details,
it is necessary to know the dark to luminous mass ratio
in present day galaxies (Dalcanton et al. \cite{dalcanton97}; 
Mo et al. \cite{mo97}). In addition
a theoretical basis can then be provided for the Tully-Fisher
relation (Tully \& Fisher \cite{tully77}; Rhee \cite{rhee96a}a, 
\cite{rhee96b}b).

Already in \cite{ostriker73} Ostriker \& Peebles 
showed that a cold isolated disc
is highly unstable to bar formation. More stable discs can be
made by putting them in a large amount of spherical (dark) matter
or when discs are hot. The observations of the stellar velocity dispersions
show that the latter is not valid. Consequently dark matter is needed
to stabilize discs, but how much? Numerical calculations are well
suited to investigate this problem. Simulations by Efstathiou et al. 
(\cite{efstathiou82}),
Bottema \& Gerritsen (\cite{bottema97b}), and Syer et al. (\cite{syer97}) 
show that the
mass contribution of a stable disc must be well below the so called
maximum disc hypothesis (van Albada \& Sancisi \cite{albada86}) 
situation. This is in
perfect agreement with the results of the stellar dispersion measurements.
On the other hand, stabilization of discs may be taken over by
a massive central bulge. Can a bulge maybe be considered as a
luminous extension of a dark halo?

Galactic bulges have a global isothermal nature in the sense
that the observed velocity dispersion is rather
uniform over the bulge's extent. Bulges
can be described consistently with models of isotropic oblate spheroids
(Kormendy \& Illingworth \cite{kormendy82}; Jarvis \& 
Freeman \cite{jarvis85}), or in other words
as rotationally flattened isothermal spheres.
 
Galaxies with morphological type ranging from Sa to Sb are equipped
with both, a disc and a bulge, and consequently there is a region
where these will overlap.
One may expect that the different kinematical properties of bulge
and disc, as observed where one or the other is dominant, are maintained
when both are present. This was suggested by absorption line observations
of the Sb galaxy NGC 2613 (Bottema \cite{bottema89}). This galaxy has a
bulge which appears to be somewhat displaced from the centre and
on one side clear double peaked line profiles are observed reminiscent of
bulge and disc having a different rotation. Such a scenario has
also been suggested by Kuijken \& Merrifield (\cite{kuijken93b}, 
hereafter KM93) to
explain the asymmetric absorption line profiles of the galaxy UGC 12591.
An asymmetric profile can be made as a superposition of a low dispersion
rapidly rotating disc and a higher dispersion more slowly rotating bulge;
KM93 show in a quantitative way that this can explain the UGC 12591 
observations. To investigate the stellar kinematics in the bulge-disc
transition region I obtained absorption line spectra of the Sb
spiral NGC 7331. The data were lying of the shelve for a while when
the discovery of a counterrotating bulge of NGC 7331
was announced (Prada et al. \cite{prada96}, hereafter P96). This
renewed my interest in this galaxy and motivated the examination
of the data.

Since the first detection of a counterrotating component in an
elliptical galaxy (Franx \& Illingworth \cite{franx88}) a frantic
search for more counterrotating galaxies started in the
astronomical community. It appeared that counterrotation
in elliptical galaxies is not uncommon which, in fact, is
not unexpected considering the likely merger history of ellipticals.
In later type galaxies, however, counterrotation is rare; known
to date are NGC 4550, of type E7/S0 (Rix et al. \cite{rix92b}),
NGC 4826 of type Sab (Braun et al. \cite{braun92}; 
Rix et al. \cite{rix95}) and NGC 4138 of type
Sa (Jore et al. \cite{jore96}).
Also claimed to possess a counterrotating component is the
Sb galaxy NGC 7217 (Kuijken \cite{kuijken93a}; 
Merrifield \& Kuijken \cite{merrifield94})
although inspection of the line profiles shows that a kinematic asymmetry
by a bulge/disc situation is more likely (see also the discussion
in the remainder of this paper).
P96 find a counterrotating component in the Sb galaxy NGC 7331,
at the 15\% intensity level, counterrotating with 50 \kmss and this
component is dynamically cold. That poses a problem; how can two
cold components exist together, one rotating with $\sim$ 250 \kmss
(the disc) and the other with 50 \kms, in the same potential?
A more detailed analysis is
certainly warranted.
In addition to the kinematics in the transition region also the
velocity dispersions of the disc and the bulge can be determined.
From this the mass-to-light ratios 
of the components can be calculated and be compared with each other.

\begin{figure}
\resizebox{\hsize}{!}{\includegraphics{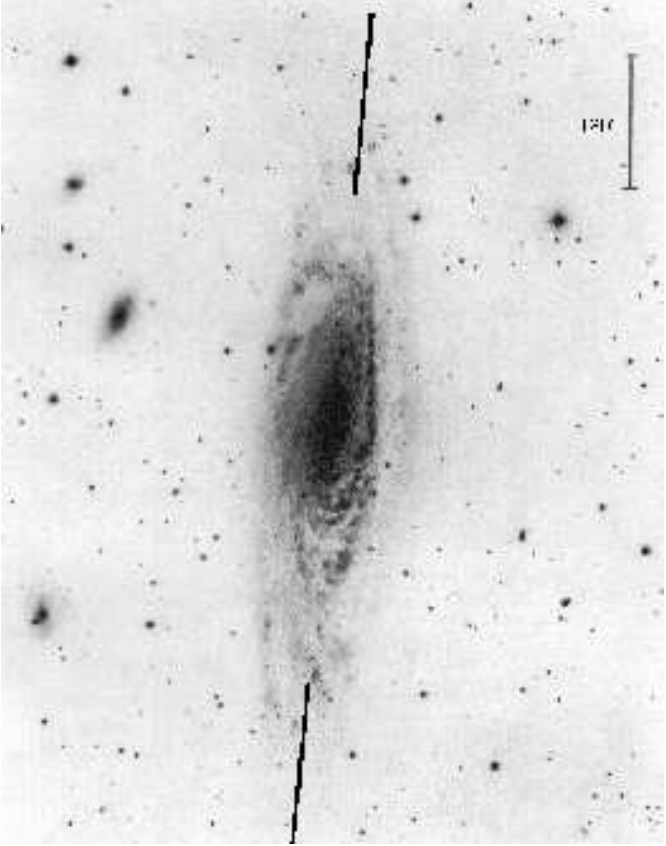}}
\caption[]{
Reproduction of the optical image of NGC 7331 from the Shapley-Ames
catalog (Sandage \& Tammann \cite{sandage81}). Spectroscopy is obtained
along the whole major axis
at the position of the left out part of the indicated line.
North at top, East at left.
}
\end{figure}

NGC 7331 is a massive Sb spiral galaxy. Photometry exists
in the r-band (Kent \cite{kent87}) and \hi observations have been
obtained and analysed by Begeman (\cite{begeman87}) and Begeman et al. 
(\cite{begeman91}).
The neutral hydrogen shows a regular velocity field and rotation
speeds up to 257 \kms. A photograph of the galaxy is given
in Fig. 1, where NGC 7331 appears to be a regular and undisturbed
system. The adopted distance of 14.9 Mpc is in perfect agreement
with the value of 15.1 $\pm$ 1.0 Mpc determined recently from
Cepheid variables (Hughes et al. \cite{hughes98}).
For convenience a summary of the main parameters is
presented in Table 1.

\begin{table}
\caption[]{Description of NGC 7331}
\begin{flushleft}
\begin{tabular}{lll}
\hline
\noalign{\smallskip}
R.A. (1950) & 22$^{\rm h}$ 34$^{\rm m}$ 47$^{\rm s}$ & $a$ \\
Declination (1950) & 34\degr\ 09\arcmin\ 30\arcsec\ & $a$ \\
Hubble type & Sb(rs)I-II & $a$ \\
Inclination & 75\degr\ & $b$ \\
Pos. angle major axis & 172\degr\ & $b$ \\
Total \hi mass & 11.3 10$^9$ $M_{\odot}$ & $b$ \\
Systemic velocity & 820 $\pm$ 3 km s$^{-1}$ & $b$ \\
Abs blue magn. $B_T^{o,i}$ & -21.73 & $c$ \\
Distance & 14.9 Mpc & adopted \\
Unit conversion & 1\arcsec\ = 72.24 pc & \\
\noalign{\smallskip}
\hline
\noalign{\smallskip}
\multicolumn{3}{l}{\quad $a$ Sandage \& Tammann (1981)} \\
\multicolumn{3}{l}{\quad $b$ Begeman (1987)} \\
\multicolumn{3}{l}{\quad $c$ from Sandage \& Tammann, for adopted dist.} \\
\end{tabular}
\end{flushleft}
\end{table}

\section{Observations and data reduction}
\subsection{The observations}
The spectroscopic observations were carried out on September 19 and
20, 1987, at the observatorio de Roque de los Muchachos on the island
of La Palma, using the Isaac Newton (2.5 m) telescope.
The intermediate dispersion spectrograph was used at the Cassegrain focus.
This spectrograph had
a grating of 1200 groves per mm, a camera with a focal length
of 235 mm (Wynne \cite{wynne77}) and slit width of 220 $\mu$m (= {1}\farcs{19}
on the sky). In that way a spectrum is produced
with a full width at half maximum
(FWHM) velocity resolution of $\sim$ 70 \kms, centred at a wavelength 
of 5162 \AA. The data were recorded with the Image Photon Counting 
System (IPCS), having a wavelength range from 4634 to 5691 \AA,
divided up into 2040 pixels. The spatial
extent on the sky was covered by 82 consecutive spectral rows each 
of size {2}\farcs{92}, resulting in a total extent of nearly four
arcminutes. During the observations the seeing was considerably smaller
than the size of a spectral row, leaving them all uncorrelated. On
both nights the slit was put parallel to the major axis, at a position
angle of 352\degr\, with the nucleus falling in the slit. One night
the northern part of the galaxy was covered with the nucleus at row
13, lying 38 arcseconds from the edge of the slit. Although in this
case also a part of the south side is covered, in the remainder of 
this paper the results from this observation will all be referred to
as ``North side''. On the second night the southern part was covered
with the nucleus being positioned {2}\farcs{3} beyond the end of the slit.
Results from this observation
will be referred to as ``South side''.

The total exposure on the North side lasted for 21\,000 seconds
(=~5~$^{\rm h}$~50~$^{\rm m}$) and on the South side for 24\,000 seconds
(=~6~$^{\rm h}$~40~$^{\rm m}$). These exposures were divided
up into single exposures of 1500 seconds, preceded and succeeded by
the observation of a Cu-Ne calibration lamp. In addition spectra of 
7 template stars all of type close to K0III were recorded, of course
using exactly the same instrumental configuration and wavelength 
calibration. Furthermore observations of the twilight sky and 
dark current were made. 

\subsection{Data handling}
For all the individual 2-d spectral images of the galaxy, the
encompassing Cu-Ne lamp exposures were added. The wavelength calibration
was performed by using 22 emission lines spaced over the 1057 \AA\
interval. The standard deviation of the lines from the final wavelength 
solution was never larger than four to five km per second. Next the data
were regridded to a common log $\lambda$ scale (30.2 \kmss per pixel)
and added to form two, North side and South side, calibrated spectral images.
For the template star exposures the same procedure was followed.
The calibrated stellar spectra were shifted to one common redshift
and after this, all spectra were averaged to produce one essentially
noise free template spectrum.

The flatfield exposures registered on the twilight sky showed that the
pixel-to-pixel variation was less than 2\%, being so small that no 
corrections were necessary. The twilight exposures were averaged together
and along the wavelength direction, providing the response along the slit.
The dark current subtracted images were corrected for this response.
There was no detectable S-shape distortion and consequently no 
corrections to remove such a distortion were necessary.
To get an impression of the count levels of the two galaxy spectra, in
Fig. 2 the average number of counts over the wavelength range is 
shown along the slit. 

\begin{figure}
\resizebox{\hsize}{!}{\includegraphics{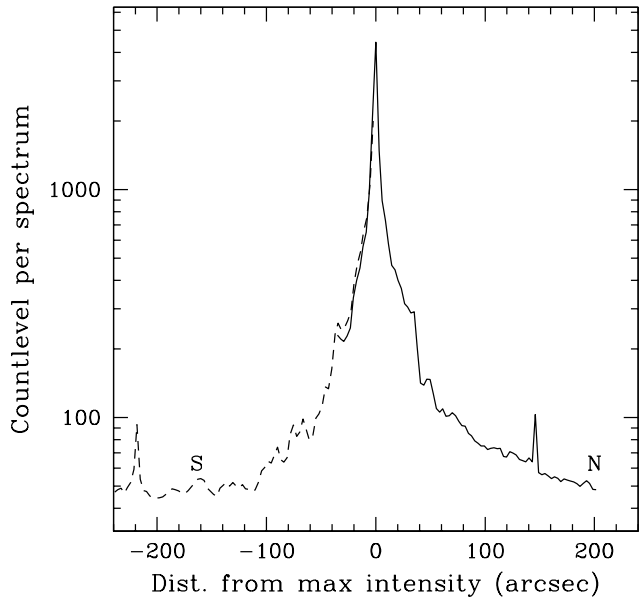}}
\caption[]{
The count level along the spectrograph slit. Given is the average
over 2040 wavelength pixels per {2}\farcs{92} on the sky.
The light of the sky is not subtracted.
}
\end{figure}

The H$\beta$ and [O~III] 5007 \AA\ and 4959 \AA\ lines were clearly
visible in the spectra. The forbidden oxygen lines continued all the way
to the centre of the galaxy, while H$\beta$ emission was essentially lost
in the H$\beta$ absorption for radii less than $\sim$ 20 arcseconds. The
emission line kinematics was determined by fitting gaussians to the profiles.
When necessary spectra were averaged along the slit to increase the S/N.
Also for the absorption lines spectra had to be averaged along the slit to
obtain a sufficient S/N.
Each spectrum has a certain position along the slit corresponding
to a fixed position on the galaxy or sky. At every position
of every spectrum, encompassing
spectra were added until a count 
level of galaxy emission of at least 
200 was reached. In this way, for larger distances from the centre
spectra become correlated but remain evenly distributed with separation
of {2}\farcs{92}. Furthermore the S/N decreases at larger radii
because of the constant level of sky and dark current emission.
When averaged in this way, the resulting spectrum is weighted
according to the intensity level of the individual spectra; the
position along the slit is (except in Fig. 4) then also given
as the intensity weighted position.

Next the continuum level was determined by fitting a fifth order polynomial 
to the whole spectrum and was subsequently subtracted. 
In this way one loses the
line strength information but the spectrum will not suffer from division by 
a number close to zero. At positions of the emission lines and strong
skylines, spectra were put at a zero level such that those positions
will not contribute to the cross-correlation sum. Certainly on the
North side and to a lesser extent on the South side there is not
a sufficient slit section free of galaxy light. Hence it is not possible
to obtain a clean sky spectrum and consequently the sky was not 
subtracted. For the present case this does not pose a large problem. 
At first because all the radial velocities of NGC 7331 are
substantially redwards from the solar radial velocity, and
no blending of the line profiles of galaxy and the sun (=sky) can 
take place. Secondly the amount of light from the galaxy is still large
enough compared to that of the sky, that a continuum level sufficiently
appropriate for the galaxy light can be determined.
As for the galaxy, from the template spectrum the continuum level
was subtracted.

The cross correlation functions have been calculated between the
spectral rows of the galaxy and the template spectrum. This is done
by using Fourier techniques. In general the continuum subtraction
by means of the polynomial fit is not sufficient and additional
low frequency wavenumbers have to be, and in the present case
were, filtered out. Insufficient filtering shows up immediately as a 
continuum level offset between the cross correlation function and
the auto correlation function of the template. Care has been taken
that such an offset did not appear. Except for the H$\beta$ absorption
line and small regions where [O~III] and skylines were present,
the whole spectral range was included in the cross correlation sum.
This range contains a forest of hundreds of absorption lines of which
the most striking are the Mg triplet around 5170 \AA\ and the
Fe/Ca complex around 5250 \AA.

\subsection{Determining the stellar kinematics: the cross-correlation-clean
technique}
Until this point the data handling is rather standard, but for getting
out the actual stellar kinematics a number of different methods
have been and are being used. All assume that the observed galaxy spectrum
can be represented by that of a suitable template spectrum convolved with
the galaxy broadening function, also called galaxy line profile.
Thus to obtain this line profile always some kind of deconvolution
is involved. The first methods to be used are those by Illingworth \&
Freeman (\cite{illingworth74}) and by 
Sargent et al. (\cite{sargent77}), applying a division
in Fourier space, and the one by Tonry \& Davis (\cite{tonry79}), applying
a cross-correlation technique. Later on other and improved methods have
been used (a.o. Bender \cite{bender90}; Rix \& White \cite{rix92a}; 
Franx \& Illingworth \cite{franx88};
van der Marel \& Franx \cite{marel93}; KM93; Statler \cite{statler95}) 
establishing a situation
where every author(s) uses his own method. In general all such
methods are being applied to elliptical galaxies or bulges, where
one has a nearly constant velocity dispersion and a large surface
brightness. For galactic discs, however, the surface brightness is lower
and the velocity dispersion decreases towards larger radii. In such
situations Bottema (\cite{bottema88}) has 
used a method which remains as closely
as possible to the original data. One can show that the cross-correlation
function (hereafter: ccf) of template and galaxy spectra is equal
to the convolution of the galactic line profile with the 
auto-correlation function (hereafter: acf) of the template spectrum
(Bottema et al. \cite{bottema87}). 
Displaying the ccf then gives one directly the line
profile, albeit convolved with a symmetric acf. By fitting a grid of 
acfs broadened to different dispersions Bottema (\cite{bottema88}) extracted
the relevant radial velocities and velocity dispersions. This method
is straightforward and the fitting can be inspected graphically, but
the error analysis had to be done by eye and asymmetric line profiles
could not be accommodated for.

Therefore this method has been improved.
In order to extract the line profile, the ccf is deconvolved using
the CLEAN technique (H\"ogbom \cite{hogbom74}), with the template acf as beam
or point spread function (see also Franx \& Illingworth \cite{franx88}). 
Certainly for
barely resolved signals with substantial amounts of noise added such a
cleaning technique proves to be a superior deconvolution method. Hence it
can be applied with confidence for the narrow, noisy, line profiles
typical for galactic discs. Clean components have been subtracted
for a region around the peak of the ccf until a residual well below
the noise level. The line profile was then restored by convolving the
components to a gaussian function with a dispersion of $2{{1}\over{2}}$
velocity pixels and by adding the residuals. The width
of this gaussian was determined as a compromise between more resolution
and more noise for a narrower gaussian and less resolution with less 
noise for a broader function. In theory, if there is no noise, one
could make the restoring function as narrow as allowed by the sampling
criterion; down to a dispersion of approximately one pixel.

When a deconvolved line profile is obtained it can in principle be 
parameterized in any way. Presently is was chosen to fit in first instance a
gaussian to the profile, even if it is not symmetric. The radial velocity
is given by the position and the stellar velocity dispersion by the
gaussian dispersion from which the (restored) resolution has been
subtracted quadratically. Errors of the parameters are obtained from 
the usual least squares fitting procedure. By fitting such a single gaussian
to the profile the position and dispersion are extracted in a way
similar to that of the conventional methods, allowing and easy
comparison and understanding.
In the case of NGC 7331 skewed profiles are observed
and consequently more parameters are needed to describe the profile.
Well suited appeared to be the Hermite polynomial analysis by
van der Marel \& Franx (\cite{marel93}). Only one additional asymmetric ($h3$)
parameter was considered, resulting in a least squares fitting of
the function

\begin{eqnarray}
f(v) &=& a {\rm e}^{{-(v-b)^2}\over{2c^2}} 
\biggl\{ 1 + h3\biggl[ 1.1547 \left( {{v-b}\over{c}} \right)^3 
\nonumber \\
 &-& 1.1732 \left( {{(v-b)}\over{c}} \right) \biggr] \biggr\},
\end{eqnarray}
\begin{figure*}
\resizebox{12cm}{!}{\includegraphics{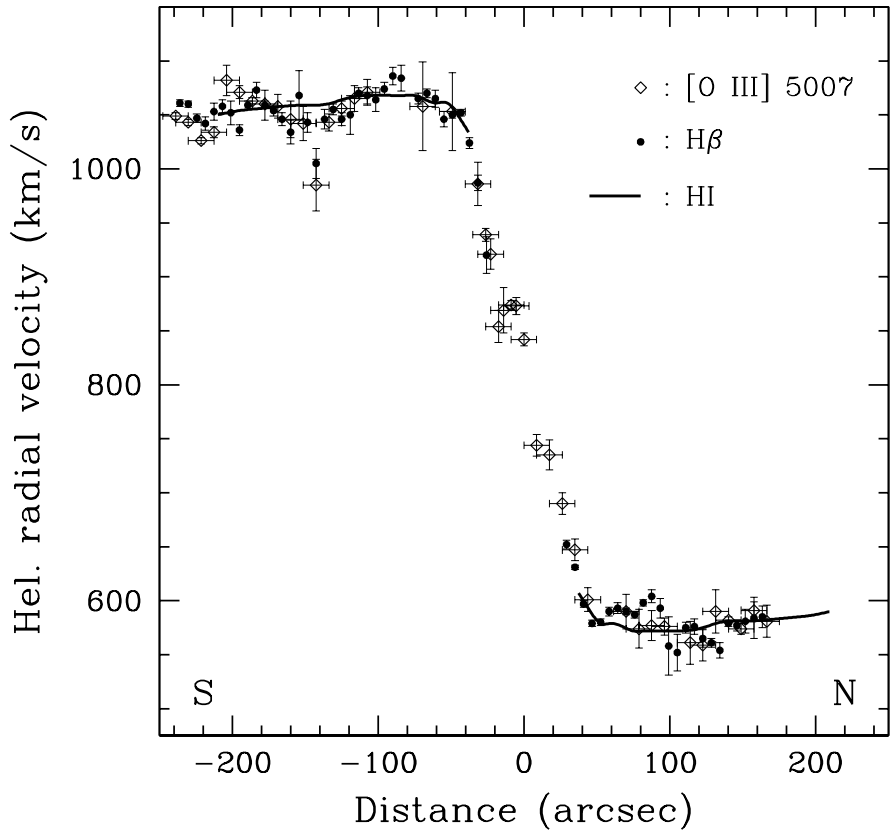}}
\hfill
\parbox[b]{55mm}{
\caption[]{
Observed radial velocities of the H$\beta$ and [O~III] 5007 \AA\ emission
lines compared with the radial velocities inferred from
the \hi rotation curve. In the inner 40\arcsec\ there is no neutral
hydrogen gas and only very little H$\beta$ emission.
}
}
\end{figure*}
to the line profile. Here $v$ is the radial velocity and $a,b,c,$
and $h3$ the parameters to be determined. There are two matters to 
be noted. At first, the values of $a,b,$ and $c$ are not completely,
but nearly equal to the parameters obtained from a straight fit
of a gaussian to the profile (see van der Marel \& Franx for a discussion).
In practice, to the line profile both, a gaussian and the 
function given by Eq. (1) is fitted. But
when results are presented $a,b,$ and $c$ are not given, only the 
gaussian fit parameters. Secondly, $h3$ does not give the skewness
of the actual line profile, but the skewness of the
resolution broadened profile. When analysing the observations
this should be taken into account.

\section{Observational results}
\subsection{The emission lines}
The radial velocities determined from the H$\beta$ and [O~III] 5007~\AA\
line are presented in Fig. 3, together with the radial velocities
inferred from the \hi rotation curve (Begeman \cite{begeman87}; 
Begeman et al. \cite{begeman91}).
As can be seen, the velocities of the three emission lines are in
excellent agreement. In the central region, for $R$ $\la$ 40\arcsec\ there is no
\hi gas and consequently no \hi rotation curve. H$\beta$ is observed
somewhat further inwards until approximately 30\arcsec. On the other hand the
oxygen line (although weak) continues all the way
to the centre. In general bulges do not show any significant
emission lines. For NGC 7331 this indicates that the disc, or some other
young galactic component is present in the centre,
embedded in the bulge. To check whether the oxygen emission indeed
originates from a cold component and not, for instance, from
bulge planetary nebulae, the [O~III] velocity dispersion was measured.
Only at the very centre, for radii less than five arcseconds this line
is broadened to 100 $\pm$ 20 \kms. Outside this radius the line is
unresolved meaning a dispersion of less than $\sim$ 30 \kms.

A comparison was made with the radial velocities along the major axis
of the H$\alpha$ and [N~II] line
as measured by Afanasiev et al. (\cite{afanasiev89}). 
Their data are confined to within 
60 arcseconds from the centre, but have a much better spatial sampling 
than for the present observation. Except for the usual absolute calibration 
error the velocities are in exact agreement, for instance the velocity
feature at -18\arcsec\ is present in both observations. A comparison
with the observations of Rubin et al. (\cite{rubin65}) 
also shows agreement, although
a detailed comparison is not possible because slit positions do not
coincide completely. 

\subsection{The stellar, absorption line kinematics}
To illustrate the appearance and quality of the absorption line
kinematics all the cleaned line profiles along the major axis
are presented in Fig. 4. This image is constructed
using the averaging procedure along the slit as discussed in the 
previous section. Therefore the indicated distances from 
maximum intensity of the individual line profiles are the distances
of the spectra around which regions have been averaged, in an
intensity weighted way. So these distances may differ slightly from
the actual distance to the centre. North and South side have been
combined on integer spectra while in reality there is a 0.2 spectral
size (of {2}\farcs{92}) shift. But these small approximations will not
generate any significant deviation from the
actual appearance of the position velocity diagram.

\begin{figure}
\resizebox{\hsize}{!}{\includegraphics{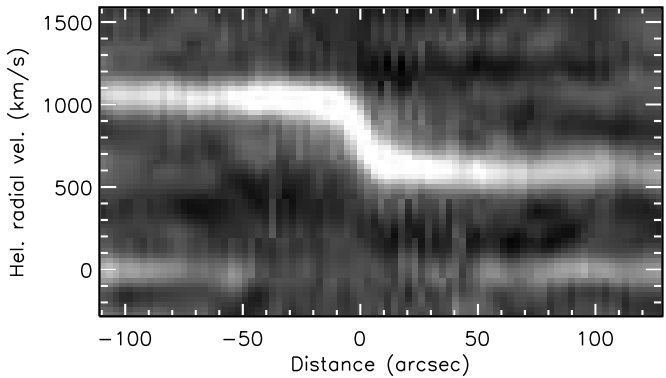}}
\caption[]{
The stellar absorption line profiles along the major axis.
For distances within 40\arcsec\ from the centre, the line profiles are skewed;
there is a shallow extension towards the systemic velocity.
Note the signature of the sky at zero radial velocity.
}
\end{figure}

The autocorrelation function of the template exhibits some corrugation
along the velocity direction. Most of this corrugation has been
cleaned away but some residual remains as can be seen in Fig. 4. In fact,
this is to be expected because the template spectrum can never be
a perfect match to the spectrum of the stellar population of the galaxy. 
Any absorption line interpretation procedure suffers from this
shortcoming, but this can only be seen by displaying matters as in 
Fig. 4. Looking at the figure one can thus notice a small gully at the
high and low end of the galaxy's radial velocities. Also some
vague reflection patterns are apparent. The magnitude of these 
irregularities is small, however, always below the noise level 
determined from line free regions surrounding the galactic emission.
In addition one may note the signal of the solar spectrum at a radial
velocity of zero.

\begin{figure*}
\resizebox{12cm}{!}{\includegraphics{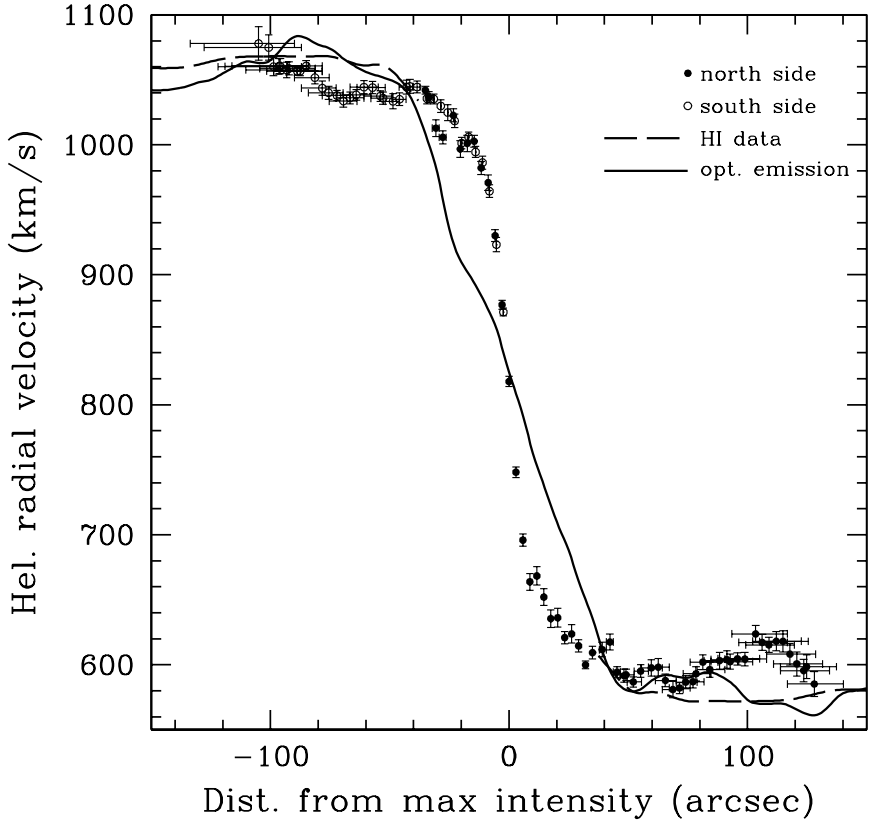}}
\hfill
\parbox[b]{55mm}{
\caption{
Stellar radial velocities compared with the emission line velocities.
The full drawn line is a fit by eye to the H$\beta$ and [O~III] data
of Fig. 3. Contrary to what is expected, in the inner regions,
the gas appears to rotate slower than the stars.
}}
\end{figure*}

By fitting a gaussian to the line profiles the stellar radial velocities
and dispersions are determined. The first are presented in Fig. 5, 
together with the \hi radial velocities and emission line
velocities, the latter given by a hand drawn line through the data points 
of Fig. 3. The velocity dispersions are presented in Fig. 6, top panel.
At the centre dispersion values around 130 \kmss are reached,
while for larger radii the dispersion decreases, similar for both
sides of the galaxy. In the lower panel of Fig. 6 the $h3$ parameter is
displayed following from a fit of Eq. (1) to the profiles. Like the
dispersion this asymmetry parameter is symmetric and regular
with respect to the centre of the galaxy.

\subsection{Discussion of the observed kinematics}
From Fig. 5 it will be immediately obvious that for radii $\la$ 40\arcsec\ the
emission line gas appears to rotate slower than the stars. This is
puzzling. One would expect the contrary; caused by asymmetric drift
and integration effects a regular galaxy shows, in general, a stellar rotation
lower than that of the gas. That is to say, the stars rotate
slower than the ``potential'' rotation or testparticle rotation
(= rotation of a testparticle in the plane: $v_t = \sqrt{-R\; 
\partial \Phi / \partial R }$). For a galaxy the emission line gas
should rotate as testparticles and consequently for NGC 7331 the
velocity aberration of this emission line gas is even larger than the
observed velocity difference between stars and gas. 
For example
at a radius of ten arcseconds the gas rotates more then 107 \kmss
too slow, which amounts to nearly a factor of two.
How can this be explained?

One might imagine a barred situation. It is known that stars
(Miller \& Smith \cite{miller79}; Hohl \& Zang \cite{hohl79}) and gas 
(Athanassoula \cite{athanassoula92}, and references therein) have
substantial non circular streaming motions along the bar. 
But at first it is hard to imagine how the gas motion might
differ so substantially from the stellar motion. For instance the analysis by
Athannasoula (1992) shows that the gas motion in a bar largely follows
the $x_1$ and $x_2$ stellar orbits. 
Secondly the I and K band photometry of the central regions
by P96 does not indicate any significant
barred morphology. Hence the explanation by a bar is unlikely.

Another possibility is that the gas ends, going inwards,
in an edge-on gas ring at $R \sim$ 50\arcsec.
Such a situation will give the kinematical appearance of
a solid body rotator as observed inside this radius. However,
given the need for a sudden inclination change from 75 to 90
degrees combined with the need for the gas to end at the same
position, this explanation seems remote.

A slower apparent (gas) rotation can, of course, be produced by a more
face-on orientation of the plane in which the gas is rotating.
To produce the difference between gas and stellar velocities as observed,
it is then necessary that from approximately 40\arcsec\ inwards the gas plane
is increasingly more inclined with respect to the plane of the galaxy. 
Such an inner warped structure could
have been generated by a gascloud or gas rich dwarf galaxy
which has fallen in on an orbit with some angle with respect to the
plane of the galaxy. Numerical calculations show that
certainly the gas will then quickly settle into the centre of the galaxy. 
This explanation is a theory which works in principle. However, to
prove or disprove it will not be straightforward; observations of
the emission line gas other than along the major axis are needed. 
A different observation which supports the hypothesis of a warped gas layer
is that of the inclined gas ring or disc inside the bulge of
M31 (Boulesteix et al. \cite{boulesteix87}), 
a galaxy similar in appearance as NGC 7331. 

Another point  to be noted is the irregular behaviour along the slit
of both the stellar
radial velocities and dispersions. For example the stellar rotation
is different 
on both sides near a radius of 100\arcsec. Also at those positions the
difference between gas and stellar rotation is different for both
sides and in addition the velocity dispersion suddenly increases over there.
This can only mean that in some regions of this galaxy stars and gas
have not yet formed a relaxed configuration. Is this kinematical appearance
of NGC 7331 unusual? No, it is not. The stellar kinematics of large
galaxies is in general more irregular than that of smaller galaxies.
These matters are discussed by 
Bottema (\cite{bottema93}) arguing
that the irregularity could be the result of more or less recent
capture of small matter clumps by large galaxies. 

\subsection{No counterrotation}
Both, the image of the line profiles in Fig. 4, and the $h3$ parameter
in Fig. 6 show that for radii less than 40\arcsec\ the line profiles
are asymmetric; they have a regular and shallow extension towards
the systemic velocity. Using a modified version of the
Unresolved Gaussian Decomposition method (KM93)
P96 find for the same object a double line profile interpreted by them
as a counterrotating bulge. This double
profile is found between radii of 5 to 20 arcseconds on both sides along
the major axis. There is one dominant component at the normal galactic
radial velocities. Another component at a 15\% intensity level of the
dominant peak is counterrotating with approximately 60 \kmss
and is unresolved by their spectrographic setup, meaning a velocity dispersion
less than $\sim$ 50 \kms. Compared to the bulge dispersion of 125 \kmss
this component is cold. A close inspection of P96's profiles shows
that although the counterrotating component is barely above the 
$1\sigma$ noise level, it is present in a number of independent
spectral rows.

In this study such a counterrotating component is not observed.
Then we have the uncomfortable situation that different analyses
of the same object give different results. How is that possible?
To check if some kind of error has been made using the present
method of interpreting the data a close inspection was made
of the original ccfns. The same asymmetry at radii $<$ 40\arcsec\
is obvious but there is no extra component. Hence the clean procedure
did not make appear or disappear any features. Because the 
cross-correlation procedure gives you directly the line profiles
it is not likely that such a counterrotating component really exists.
Note that P96 have nearly the same instrumental resolution, but
spectra are taken around 8500 \AA\ instead of the present 5100 \AA.
Could P96 have done something wrong? To
investigate this the UGD algorithm has been applied to the present
North side spectra. The same UGD setup with a two pixel dispersion
and three pixel separation of the gaussians is used
as by P96, and with nearly equal velocity pixel sizes both
analyses should be comparable. And the result? Also a counterrotating 
separate line profile is found at the 20\% level of the main profile
between radii of 10 and 40\arcsec\, counterrotating with 50 \kms.

The analysis of P96 differs slightly from the
standard UGD algorithm. They use a two dimensional
version of UGD where just as in the dispersion direction, also
in the spatial direction gaussian components are fitted with a
dispersion of 2 pixels and separation of 3 pixels. This
imposes extra smoothness and noise suppression 
(Binney \& Merrifield \cite{binney98}).
But, frankly, I do not see that such a procedure is much
different from a simple smoothing along the spatial direction,
giving better S/N and poorer resolution. P96 point out that:
``in general, the results of both algorithms (UGD and 2dUGD) are
the same''. As demonstrated above, the original UGD
algorithm indeed gives the same result as found by P96 and consequently
in the present discussion it should not matter whether
the 1dUGD or 2dUGD procedure is used. 

Anyway, the UGD method generates a counterrotating
component while the cross-correlation clean (CCC) method only
produces an asymmetric profile.
A more detailed comparison between the two methods has been
performed, doing some
test analyses on skewed profiles. A description of this investigation
is given in the Appendix. It appears that the counterrotating
component is most likely an aliasing feature produced by UGD.
When errors are superposed on the lineprofile as calculated by UGD,
this feature is not significant and should have been discarded.
So, UGD is not wrong, but by omitting the presentation of the
errors P96 may have been led to a wrong interpretation of the results.

\begin{figure}
\resizebox{\hsize}{!}{\includegraphics{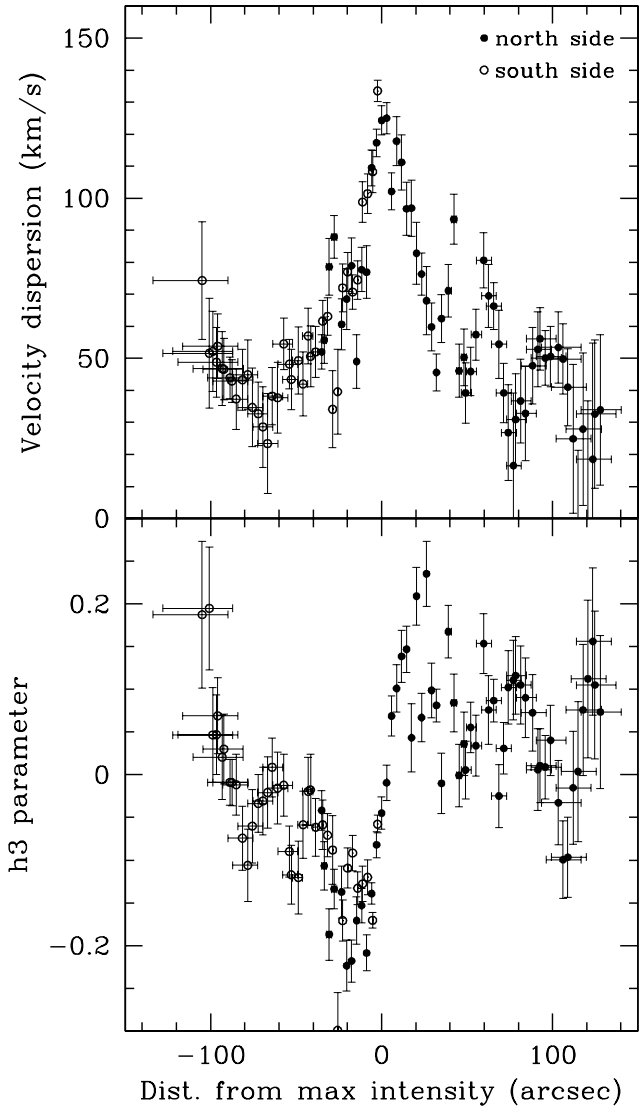}}
\caption[]{
Stellar velocity dispersions and absorption line asymmetry $(h3)$
parameter along the major axis. Both show a regular
and symmetric behaviour with respect to the centre.
}
\end{figure}

\section{The bulge/disc light decomposition}
\subsection{The observed photometry}
To interpret the observations a dynamical model of NGC 7331 will be
constructed. For this a description is necessary of the luminosity
distribution of the bulge and the disc which have to be
derived from the photometry. In the present case such a procedure
is not straightforward.

Photometry in the Thuan \& Gunn (\cite{thuan76}) r-band 
is obtained by Kent (\cite{kent87}).
Unfortunately a bulge/disc light decomposition according to Kent's
procedure is not possible because the apparent ellipticity of the 
bulge and disc are equal. Therefore one has to resort to
the classical method of assuming a certain density behaviour of one
of the components, fit this behaviour there where this component
is dominant and subtract to get the density of the other component.
For example Kent (\cite{kent87}) assumes an exponential bulge with scalelength
of approximately 0.8 kpc, which he subtracts from the total photometry
to get the disc light. In the rotation curve analysis by Begeman
et al. (\cite{begeman91}), for the same photometry a disc with scalelength
of 62\arcsec\ has been fitted to the outer regions. Subtraction
of the disc gives the bulge light and a bulge to disc light ratio
of $\sim$ 1.4, which is too large for this galaxy.

Photometry of the inner regions in the I and K bands is presented
by P96. Although P96 argue for a different radial behaviour in
K and I, in reality, overlaying the two profiles shows that both are 
nearly identical over the radial extent where both have been measured.
In the remainder of this paper the I band photometry of P96 will be used.

The r-band observations show a constant ellipticity
$\varepsilon$ of 0.6 for the whole galaxy. In the I and K band, however,
the central regions show a tendency towards a rounder situation.
Fig. 2 of P96 gives the ellipticity along the major axis. For radii
larger than 15\arcsec\ $\varepsilon = 0.6$ consistent with Kent's
photometry; going inwards, the ellipticity decreases gradually
to $\varepsilon = 0.4$ at 5\arcsec\ and to $\varepsilon \sim 0.1$ near
the centre. Apparently there is a nearly spherical distribution
very close to the centre.

\begin{table*}
\caption[]{Bulge/disc decomposition parameters}
\begin{flushleft}
\begin{tabular}{llllll}
\noalign{\smallskip}
\hline
\noalign{\smallskip}
bulge & $h_d$ & ${\mu}_0$ & $R_e$ & ${\Sigma}^m_0$ & 
$c/l_d(0,0)$ \\
name & (arcsec) & (I-mag  & (arcsec) & (I-mag 
& (Eq. 3) \\
 & & arcsec$^{-2}$) & & arcsec$^{-2}$) & \\
\noalign{\smallskip}
\hline
\noalign{\smallskip}
large & 43 & 18.7 & 37 & 12.0 & 488 \\
small & 26.5 for $R <$ 60\arcsec\ & 
17.7 & 15 & 10.9 & 432 \\
lpd & see Fig. 7 & see Fig. 7 & 39 & 12.7 & 31 \\
\noalign{\smallskip}
\hline
\end{tabular}
\end{flushleft}
\end{table*}

\subsection{The ``large'' and ``small'' bulge}
There remains the decomposition problem. Initially two extreme
bulge/disc ratios were chosen expected to encompass the range of
possibilities. These ratios or situations will be referred to
as ``large'' bulge and ``small'' bulge. The large bulge was constructed
in the following way: Stellar kinematical information is available till
$R \sim $ 150\arcsec; therefore a fit of an exponential disc was made
to the outer range of this 150\arcsec\ resulting in a disc scalelength
$(h_d)$ of 43\arcsec\ and disc central (not corrected to face-on)
surface brightness ${\mu}_0^I$ of 18.7 mag. arcsec$^{-2}$. This disc
was subtracted from the total I-band photometry and to the remainder
an $R^{1/4}$ law:

\begin{equation}
{\Sigma}^m (R) = 8.33\; (R/R_e)^{1/4} + {\Sigma}^m_0 ,
\end{equation}
was fitted, where ${\Sigma}^m$ is the surface brightness in magnitudes,
${\Sigma}^m_0 = {\Sigma}^m (R=0)$, and $R_e$ the effective radius.
In this and forthcoming cases Eq. (2) always proved to be a good
representation of the bulge light. Bulge and disc parameters are
given in Table 2.

The smallest possible bulge contribution was
determined as follows. If the bulge is spherical as indicated
by the central ellipticity and the disc generates the larger ellipticity
for $R >$ 15\arcsec\, then the bulge/disc transition should be near the
middle of this radial extent. It appeared that such a situation can be
achieved using an exponential disc with scalelength of {25}\farcs{5}
until $R =$ 60\arcsec\ and following the total light for $R >$ 60\arcsec.
A disc with this scalelength is equal to the disc fit of P96 to the
K-band data, but that is a coincidence.
Again the bulge was found by subtracting the disc
from the total light and fitting an $R^{1/4}$ law, of which the parameters
are also given in Table 2. This whole procedure constructing the large
and small bulge may seem somewhat arbitrary and in fact it is.
Still I feel confident that a good determination is given
of the range of bulge/disc ratios; which may be judged by inspecting
the graphical representation of the result given in Fig. 7.

\begin{figure}
\resizebox{\hsize}{!}{\includegraphics{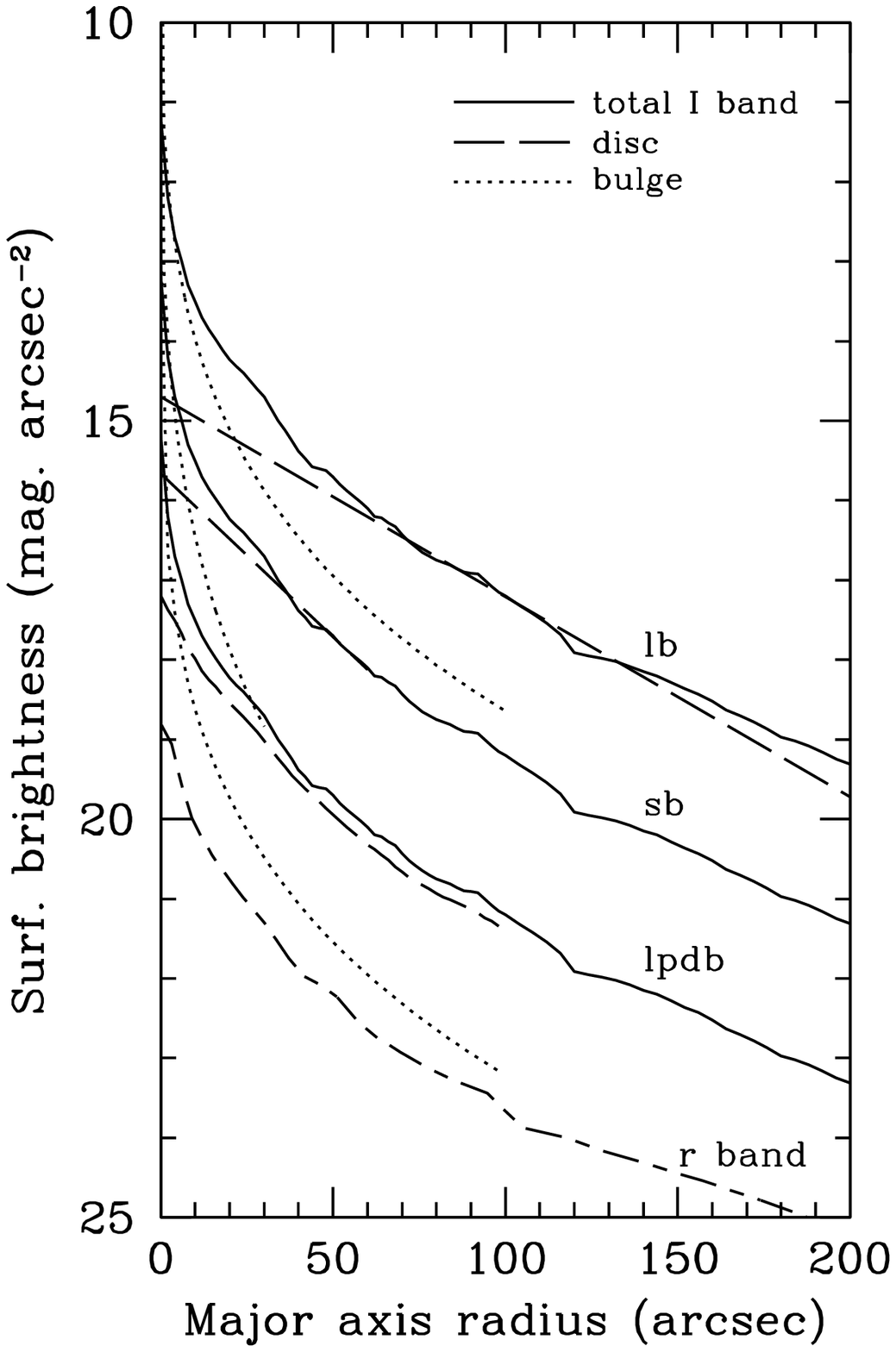}}
\caption[]{
Elliptically averaged $I$ and $r$ band photometric profiles
of NGC 7331. The lower $I$ profile is the actual one, the middle
and top $I$ profiles are offset by -2 and -4 mag. and the $r$-band
profile is offset by +2 magnitudes. Three bulge/disc
decompositions are indicated; lb for a large bulge, sb for a small
bulge, and lpdb for a line profile determined bulge situation.
In practice  the sum of bulge and disc profile is
(except for the inner 2\arcsec) indistinguishable from
the observed total light profile.
}
\end{figure}

\subsection{The ``lpd'' bulge}
Nevertheless when comparing the results of the models with
the actual data, both large and small bulge did not give
satisfactory results. Guided by experience obtained when
doing this modelling it became clear that in fact a kinematical
light decomposition could be made. That is to say, the narrow high
velocity component of the line profile has to originate from
the disc, while the broad underlying component comes from the bulge.
Therefore an estimate of the bulge/disc light ratio was made by
considering the shape of the line profiles, all as a function
of radius. Such a determination is not very precise, but the initial
resulting b/d ratio already produced a better agreement between model
results and data. Results were improved by increasing the initially
estimated global b/d ratio with a factor two. In practice, the derived
b/d ratio as a function of radius was applied to the I-band photometry
producing a disc and bulge luminosity radial profile.
Note that in this way both, the bulge and the disc radial light profile
can be determined without any a priori constraint on the radial
functionality. To the bulge profile again an $R^{1/4}$ law was 
fitted and then results a 
``line profile determined'' bulge, or from now on ``lpd'' bulge, which
is given graphically in Fig. 7. This lpd bulge has an effective radius
comparable to that of the large bulge, while its surface brightness
is lower (Table 2), which ensures that it falls between the two extremes
considered above.

\subsection{Deprojection}
To put the bulge into a 3d model for the galaxy the $R^{1/4}$
bulge(s) have to be deprojected. For this a slight variation of the
formula given by Young (\cite{young76}) is used

\begin{equation}
l_b(r) = c\; {\rm exp}(-7.67\; s^{1/4})\; s^{-7/8}
(1 + 0.11\; s^{-1/4})^{-1},
\end{equation}
with

\begin{equation}
s = {{r}\over{R_e}},
\end{equation}
where $l_b$ is the luminosity density as a function of 3d 
coordinate $r$  and $R_e$ is the observed
effective radius.
In this case the bulge is assumed
to be spherical. To
check the influence of this assumption also situations with 
a flattened bulge (with $a/b =$ major/minor apparent axis = 2/1)
have been investigated. It appeared that the
results of the stellar kinematical modelling
are not sensitive to such changes of the bulge morphology.
Bulge and disc
were put in a 3d situation and integrated along the line
of sight to determine the constant of proportionality $c/l_d(0,0)$ 
($c$ from Eq. (3) and $l_d(0,0)$ the central luminosity density
of the disc)
by comparison with the observed b/d light ratios 
along the major axis for the different
situations. The resulting value is given in Table 2.

\subsection{The total light}
For the light along the major axis decompositions have now been
described, which is the most important since the observed spectra are
also situated along this major axis. To determine the total light
of the bulge and disc, however, one has to know the b/d light ratio
for the whole image. It took a small investigation to find
a consistent picture for this galaxy.

Near the centre the ellipticity $(a/b = 1/1-\varepsilon )$ 
indicates a round situation, so at least for $R$ $\la$ 5\arcsec\
the bulge is likely spherical. Inspection of the
image in Fig. 1 shows that a hazy component, which may be associated with
the bulge, extends to at least half the image of the galaxy
and appears to be pretty round. But that is, of course, only limited
evidence that the bulge is spherical everywhere. Additional evidence
comes from considering the disc.

The velocity field following from the 21 cm emission line gives
an inclination of 75\degr\ with small error of one to two degrees.
The ellipticity a disc with an inclination of 75\degr\ would produce
is 0.72 for an intrinsic $b/a$ ratio of 0.11 (Guthrie \cite{guthrie92}). 
This differs considerably from the observed ellipticity of 0.60. 
An explanation for this difference can only be found when NGC 7331 has
a bulge which is much rounder than $\varepsilon = 0.60$, producing
that observed value in combination with an $\varepsilon = 0.72$ disc.
Combination of three pieces of evidence then leads to the 
conclusion that NGC 7331 has a spherical or nearly spherical bulge
and a disc with projected ellipticity of 0.72 as expected for an
inclination of 75\degr. Given the lightprofiles (in the I-band) along
the major axis for the three b/d decompositions the total
light of bulge and disc can then be calculated and is given in 
Table 3. Note that to obtain the bulge light the $R^{1/4}$ fit has
been integrated till a radius of 150\arcsec; because of this fitting
the total light differs slightly for the different decompositions.

\begin{table}
\caption[]{Total light of bulge and disc}
\begin{flushleft}
\begin{tabular}{llll}
\noalign{\smallskip}
\hline
\noalign{\smallskip}
Bulge type & Disc light & Bulge light & b/d ratio \\
 & $(10^9 \; L_{\odot}^I)$ & $(10^9 \; L_{\odot}^I)$ & \\
\noalign{\smallskip}\hline \noalign{\smallskip}
large & 10.1 & 16.2 & 1.6 \\
small & 13.9 & 9.2  & 0.7 \\
lpd   & 12.7 & 11.4 & 0.9 \\
\noalign{\smallskip}
\hline
\end{tabular}
\end{flushleft}
\end{table}

\section{The model calculations}
\subsection{Reconstruction of the line profile}
The observed spectroscopic data have to be related to the real
internal velocity dispersions and stellar rotation of the galaxy.
To that aim a 3d model of NGC 7331 will be employed consisting
of a disc with a prescribed density distribution and a bulge giving
an $R^{1/4}$ law in projection. The effect of a dust layer has been
neglected, as to the validity of this one is referred to the discussion
in Bottema (\cite{bottema89}). The observable dispersion
generated by the disc is mainly determined by the dispersion in the
tangential direction. For the inclination of 75\degr\ a small fraction
of the perpendicular dispersion enters into the observations and because
the disc has a certain thickness, also a fraction of the radial dispersion.

Along the line of sight an integral line profile develops. This process
is simulated by numerically integrating the kinematical model
of the galaxy. The line profile obtained in that way was then
smoothed to the resolution of the restored observations. Next
a gaussian and $h3$ parameter were fitted exactly as for the observations.
In this way results of the model calculations are
precisely comparable to the observational data.
Very close to the centre, typically for $R$ $\la$ 2\arcsec\ both, the
model calculation and the comparison with observations cannot be
done reliably. This is because
the integration step was 1\arcsec\, the seeing is of the same order and
so close to the centre the 3d density distribution of the bulge
is not well behaved.

\subsection{The disc model}
For the disc of NGC 7331 a locally isothermal distribution 
(Spitzer \cite{spitzer42})
with constant thickness has been assumed 
(van der Kruit \& Searle \cite{kruit82}),

\begin{equation}
l_d(R,z) = l_d(R,0) {\rm sech}^2 \left( {{z}\over{z_0}} \right),
\end{equation}
where $l_d$ is the luminosity space density and $z_0$ indicating the
thickness of the disc. In the radial direction the luminosity
density was proportional to the surface brightness as given
by the dashed lines in Fig. 7, $l_d(R,0) \propto {\mu}(R)$.
For all cases the value of $z_0$ has been put at 7\arcsec\ which
is a compromise between one fifth of the scalelengths
(van der Kruit \& Searle \cite{kruit82}) of small and large bulge situations.
A constant M/L ratio is adopted which gives
a vertical velocity dispersion $\dispz$ for the locally isothermal
disc of Eq. (5) of

\begin{equation}
\dispz = \sqrt{\pi G z_0 {\sigma}_0 } \sqrt{f(R)} =
\dispzn \sqrt{f(R)}, \end{equation}
with

\begin{equation}
f(R) \equiv {\mu}(R)/{\mu}_0. \end{equation}
Here ${\sigma}_0$ is the central surface density, ${\mu}(R)$ the observed
(so not corrected to face-on) surface brightness and ${\mu}_0 =
{\mu}(R=0)$. Note that

\begin{equation}
(M/L)_{\rm disc} = {\sigma}_0/{\mu}_0^{\rm face-on} . \end{equation}
The velocity dispersion in the radial direction is given by

\begin{equation}
\dispr = {{1}\over{0.6}} \dispz, \end{equation}
following from observations of old disc stars in the solar neighbourhood,
comparison of velocity dispersions of inclined and face-on galaxies
(Bottema \cite{bottema93}), and by results from numerical heating experiments 
(Villumsen \cite{villumsen85}). 
Finally the velocity dispersion in the tangential
direction follows from

\begin{equation}
{{\dispt}\over{\dispr}} = \sqrt{{{B}\over{B-A}}}, \end{equation}
where $A$ and $B$ are the Oort constants, which can be derived from
the rotation curve. The local velocity distribution is assumed
to be gaussian in the three directions.

The input stellar rotation curve $(v_{\ast})$ is given by the
testparticle rotation ($v_t$, see Sect. 3) diminished with
the asymmetric drift of the stars (van der Kruit \& Freeman \cite{kruit86};
Binney \& Tremaine \cite{binney87}). For a plane parallel disc this
drift $(v_t - v_{\ast})$ is given by

\begin{eqnarray}
v_t^2 - v_{\ast}^2 &=& \langle v^2_R \rangle
\biggl[ {{-R}\over{\rho}} {{\partial \rho}\over{\partial R}}
- {{R}\over{\langle v^2_R \rangle}} {{\partial}\over{\partial R}}
\langle v^2_R \rangle \nonumber \\
 &-& \left({{A}\over{A-B}} \right) \biggr],
\end{eqnarray}
which, using Eq. (6) and (9) is equal to

\begin{equation}
v^2_t - v^2_{\ast} = \langle v^2_R {\rangle}_{R=0} f(R)
\left[ {{-2R}\over{f(R)}} {{\partial f(R)}\over{\partial R}}
- \left( {{A}\over{A-B}} \right) \right]. \end{equation}
Usually one can take the rotation of the ionized gas to represent
the testparticle rotation. Definitely for $R$ $\la$ 40\arcsec\ this cannot
be done for NGC 7331. To resolve this problem an iterative strategy has
been followed. As a first guess, from 60\arcsec\ inwards a constant
rotational velocity of 250 \kmss was assumed. Observable radial
velocities and dispersions were calculated for a certain bulge/disc situation
and by comparison with the observations the input testparticle rotation
together with the other input parameters were adapted to find
a satisfactory fit. It proved possible to fix the input rotation quite
independently from these other parameters, to end up at
a situation where the testparticle rotation increases gently
from 255 \kmss at $R =$ 70\arcsec\ to 270 \kmss near the centre.
It will be obvious that only such a global trend can be deduced
and no details of the rotation curve.

\subsection{The bulge model}
The three dimensional density distribution of the bulge is given
by Eq. (3), and the ratio of bulge to disc
density was fixed by comparing the surface brightness calculated
from the model distributions with the actual
bulge/disc surface brightness along the major axis as given
in Fig. 7. For the bulge kinematics a simple model was assumed.
The velocity dispersion is isotropic and constant like an
isothermal sphere. The bulge is cylindrically rotating with
a rotation curve

\begin{equation}
v^{\rm b}_{\rm rot} = v^{\rm b}_{\rm max}
\sqrt{ 1 - {{d_{\rm b}}\over{r}} {\rm arctan}\left(
{{r}\over{d_{\rm b}}} \right) },\end{equation}
where the bulge rotational core radius $d_{\rm b}$ was kept fixed
at 1\arcsec. It finally appeared that the maximum bulge
rotation $v^{\rm b}_{\rm max}$ has to be considerably smaller than the
disc rotation, but its value is only loosely constrained. Consequently
results and conclusions are not sensitive to any reasonable change
of this bulge rotational core radius.

\section{Fitting the data for the ``large'' and the ``small'' bulge}
To start this section a short description will be given of the main
effects which there are on the observable quantities when
changing a certain input parameter of the galaxy model.
Near the centre the bulge always dominates the light and hence
the bulge dispersion is fixed by the dispersion observed near the
centre. In the outer regions the disc dominates the dispersion but
the bulge influence is not negligible. Over there the disc
generates a high and narrow line profile while the bulge contributes
a broad low level component to the profile. Because of the
relatively broad (compared to the disc) instrumental profile this
underlying bulge component is easily picked up even if the bulge
is weak. One may wonder whether the bulge light continues 
substantially into the disc region and whether it is not cut off
at a certain radius. The optical image of this galaxy argues against
such a proposition; the bulge remains visible as a faint haze
well until radii where the spectroscopy ends. Simulations with a disc
only did not produce detectable asymmetric line profiles.
Consequently any asymmetry or large $h3$ value must be generated
by an unequal bulge and disc distribution and/or kinematics.
Increasing the difference between the disc and bulge rotation
increases the observable dispersion in the transition region.
In the mean time the observable asymmetry then becomes larger.

First the large bulge situation was considered but it appeared impossible
to match the observations. Whatever the testparticle rotation
of the disc, the bulge rotation in combination with its light distribution
dominate the observable rotation, typically for radii $\la$ 40\arcsec.
This results in a rotation much lower than what is
observed. The dispersion was always too large
between radii of 20 to 80\arcsec\ even when the disc dispersion was
made very small. In addition, the $h3$ parameter could never be made
large enough. To summarize: a large bulge situation is really
out of the question.

\begin{figure}
\resizebox{\hsize}{!}{\includegraphics{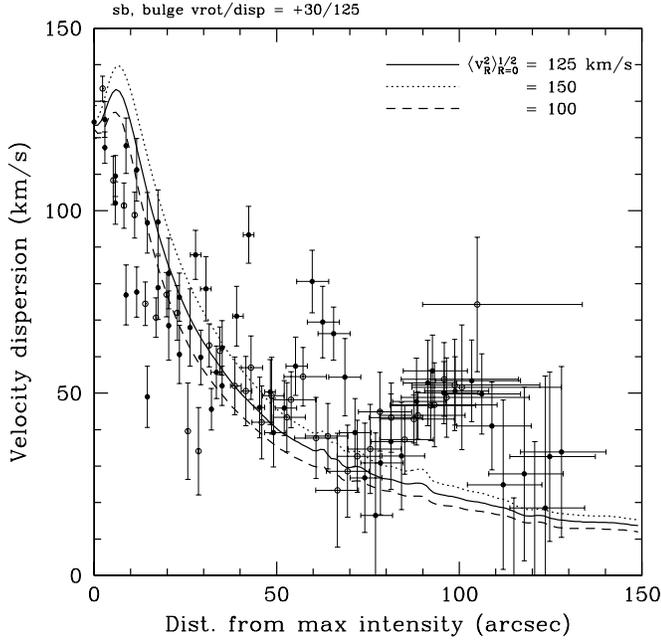}}
\caption[]{
Comparison of velocity dispersions obtained from
a model calculation (the lines) with the observed stellar velocity
dispersions. All for a small bulge light decomposition, a bulge
rotation and dispersion of 30 and 125 \kmss respectively. 
A best fit is for a disc central dispersion in the radial direction
$(\disprn)$ of 125 \kms. The fit is reasonable but not perfect.
}
\end{figure}

Next the small bulge was considered. The model parameters were adjusted
to obtain the best fit to the observations of which the result is shown
in Fig. 8 and 9 for the dispersion and asymmetry respectively.
This best fit has a central disc dispersion $\disprn$ of
125 $\pm$ 25 \kms, a bulge dispersion of 125 $\pm$ 10 \kmss and bulge
rotation of approximately 30 \kms. For a comparison of model and observed
radial velocities and determination of the input testparticle rotation
one is referred to the next section and Fig. 14, because results
concerning these matters are equal for the small and lpd bulge.
There appears to be a reasonable resemblance in figures 8 and 9
between the data and the model, but the fits are certainly not perfect.
For instance the model dispersion decreases nicely with radius,
but too steeply compared to the observations. At the smaller radii,
typically for $R$ $\la$ 30\arcsec\ the fitted dispersion is
somewhat too large while for radii $\ga$ 30\arcsec\ it is too small.
The $h3$ parameter (Fig. 9) also follows the data, but the model
values are a factor two smaller than the observations. One can make a
larger $h3$ parameter by lowering the bulge rotation to zero or to negative
values (= counterrotation). But then also the dispersion values at the 
transition region between bulge and disc at $R$ $\sim$ 10\arcsec\ increase
and that is not consistent with the observations. Already for the present best
fit situation for the small bulge the model dispersion is a bit too
large at $R \sim$ 10\arcsec\ so that one would prefer a larger bulge
rotation, but then the $h3$ parameter gets smaller. Lowering the disc
dispersion to decrease the dispersion at the transition region
gives too low dispersions at large radii and is not a solution to
the problem.

\begin{figure}
\resizebox{\hsize}{!}{\includegraphics{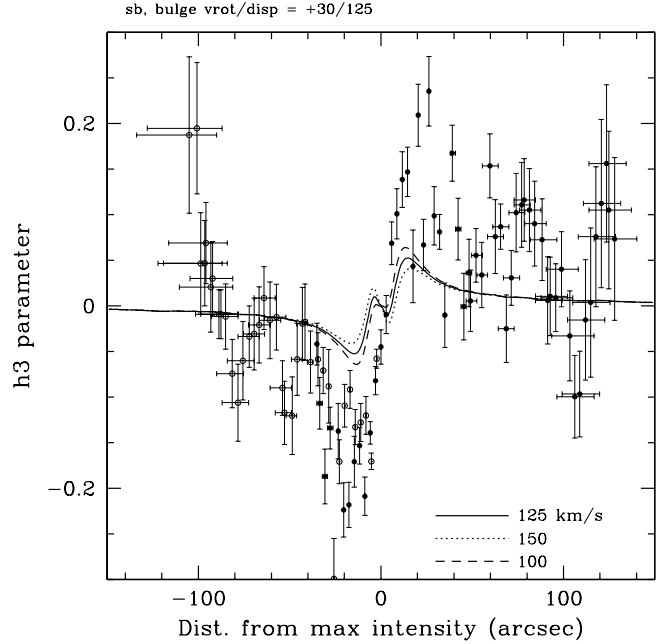}}
\caption[]{
As Fig. 8, but now for a comparison of model and observed
asymmetry parameter.
Although the radial functionalities agree, the model
asymmetry parameter is too small by a factor of two. 
}
\end{figure}

Several combinations of input parameters have been investigated, and
the final best fitting result is given in Figures 8 and 9, a result
which is not perfect and not satisfactory. The experience obtained when
doing these fits for the small bulge situation paved the way
towards a better solution. It became clear that in order to get a larger
$h3$ parameter a comparable bulge and disc intensity is needed over
a larger range of radii. The largest asymmetry is there where the
intensities are equal and the observations then indicate that this 
is between radii of 10 to 20\arcsec. Considering this, a larger effective 
radius of the bulge is needed, but the bulge should be less bright
than for the large bulge situation. Plotting of the model profiles
at diverse radii showed that the individual contributions
to this profile of bulge and disc can be distinguished and traced
as a function of radius. This knowledge was applied to the observed
profiles and it became clear that these too contain information
about relative brightness of bulge and disc, leading to the
construction of the lpd bulge situation.

\section{Fitting the data for a ``lpd'' bulge}
As in the previous section various combinations of input parameters
have been investigated. Comparison with the observations gave the best
fit which is shown in figures 10 and 11 for the dispersion and asymmetry
parameter as a function of radius. The central dispersion in the radial
direction $(\disprn)$ was determined at 140 $\pm$ 30 \kmss for a bulge
rotation and dispersion of 30 and 125 $\pm$ 10 \kms. This last value
is fixed unambiguously because the bulge dominates the light near the
centre. Although a perfect fit could not be achieved for this bulge/disc
decomposition, results are better than for the small
bulge situation. A larger asymmetry parameter is obtained more in
accordance with the observations. In addition the radial behaviour of the
dispersion follows the data reasonably well.

\begin{figure}
\resizebox{\hsize}{!}{\includegraphics{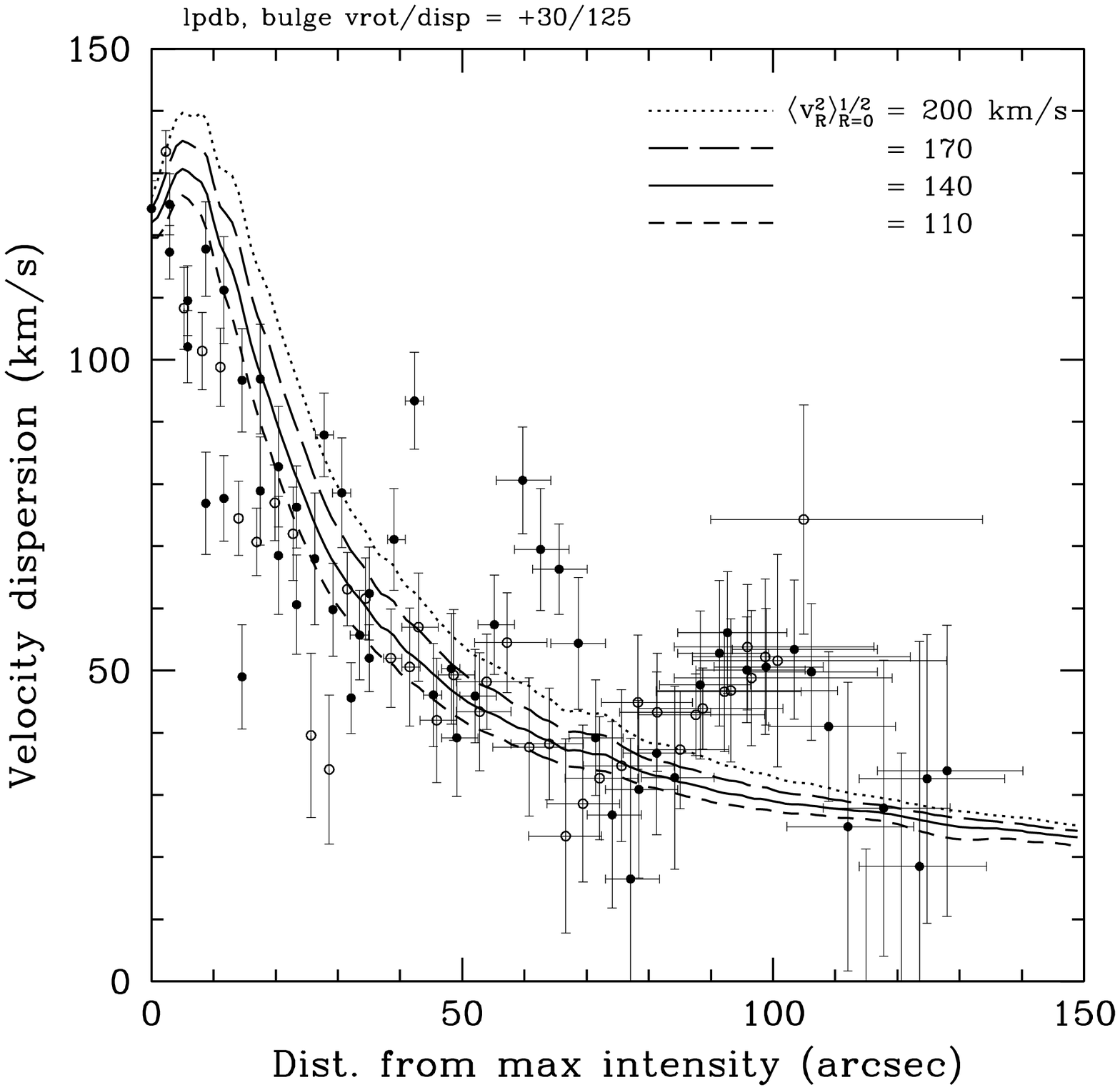}}
\caption[]{
Comparison of velocity dispersions obtained from
a model calculation with the observed velocity dispersions.
In this case for an ``lpd'' bulge/disc light decomposition
and a bulge rotation and dispersion of 30 and 125 \kms respectively.
The best fit for $R >$ 25\arcsec\ is for $\disprn = 140 \pm 30$ \kmss
and shows a better agreement with the data than for the small bulge
situation. For $R$ $\la$ 25\arcsec\ a somewhat smaller dispersion is
preferred.
}
\end{figure}

\begin{figure}
\resizebox{\hsize}{!}{\includegraphics{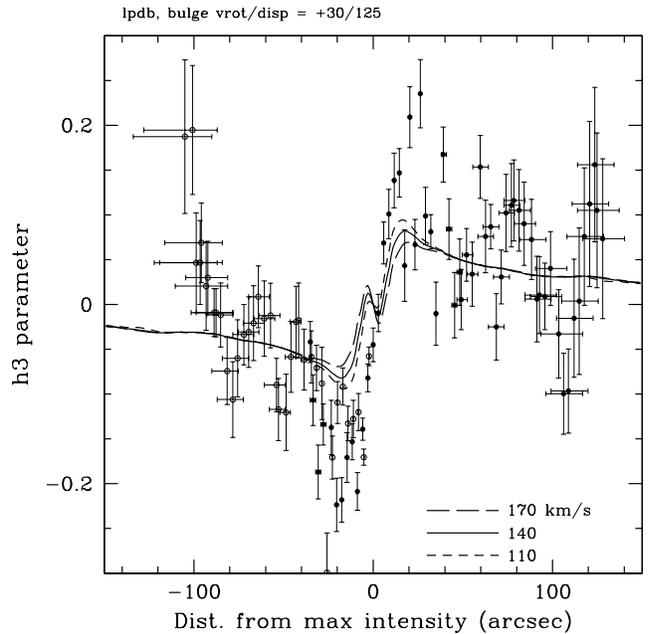}}
\caption[]{
As Fig. 10, but now for a comparison of
model and observed asymmetry parameter. Note that considerably
larger $h3$ values can be obtained than for the small bulge model.
}
\end{figure}

The rotation of the bulge is not so well constrained. Effects of a
different bulge rotation are shown in figures 12 and 13. Again
the same trends as described in the previous section are apparent;
a lower bulge rotation gives a larger $h3$ parameter
and larger dispersion at radii between 5 to 20\arcsec\ where bulge
and disc are comparable in brightness.
For a bulge counterrotating with 20 \kmss a good fit is found to the
asymmetry parameter, but the dispersions around 10\arcsec\ are 
considerably larger than what is observed. When the bulge rotates with
80 \kms, the fit to the dispersions is perfect, but $h3$ values are too
small. Therefore a compromise is found for a bulge rotation
of 30 $\pm \sim$ 30 \kms.
In this respect it should be noted
that the observed velocity dispersion is more trustworthy than the
observed $h3$ parameter, because the latter is more susceptable
to small scale irregularities. 
For instance the small gully on the high and low velocity
side of the line profiles (Fig. 4) may have caused the $h3$ values 
to be determined somewhat too large. The velocity dispersion is always
a more global parameter. 

Does the fact that the observed $h3$ parameter can only be
explained by a bulge counterrotating with 20 \kmss prove that
NGC 7331 indeed has a counterrotating bulge? If such a situation
as explained above and illustrated in Fig. 13 is valid, it is totally different
from the counterrotating bulge as proposed by P96. That bulge is cold
and counterrotates with $\sim$ 50 \kms, while the present bulge is
hot (disp = 125 \kms) and counterrotates with only 20 \kms.
Some test line profile calculations have been done for a cold
counterrotating bulge. As can be expected in the inner 5\arcsec\ the
calculated dispersion is much lower than observed and between
8\arcsec and 25\arcsec\ the dispersion is much larger than observed.
The $h3$ parameter is irregular as a function of radius and even
has the wrong sign at certain positions. 
Thus a cold counterrotating bulge cannot be made in agreement with the
observed kinematics. Instead of a completely counterrotating bulge
one might argue for a counterrotating component inside this galaxy.
This is not easy to investigate because a spatial distribution
for such a component can only be guessed. In any way,
such a component would increase the dispersion at the positions where
it is situated, which is difficult to reconcile with the observed
dispersions. Still, a small counterrotating component inside the
bulge or at the bulge/disc transition region cannot completely 
be excluded. Also judging the fits in Fig. 12 and 13 a slowly
counterrotating hot bulge might just be possible.

\begin{figure}
\resizebox{\hsize}{!}{\includegraphics{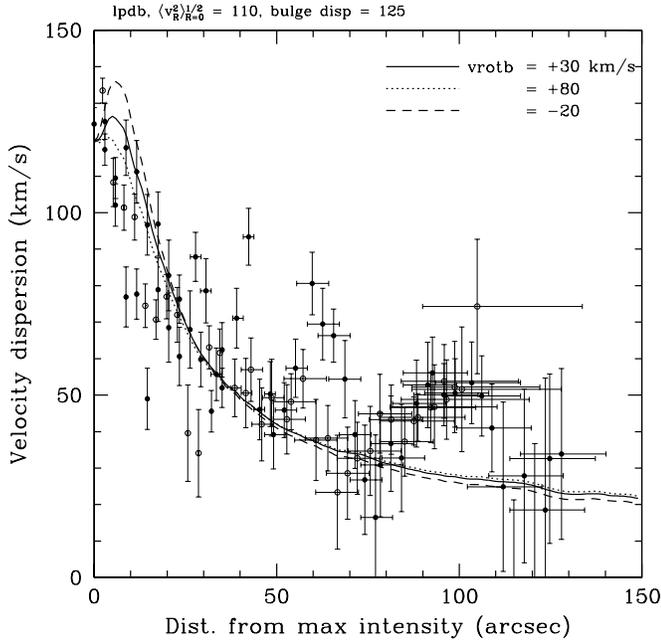}}
\caption[]{
Illustration of the effects of a different bulge rotation on the
observable dispersions. A larger difference between bulge and disc
rotation gives a larger dispersion in the bulge/disc transition region.
}
\end{figure}

\begin{figure}
\resizebox{\hsize}{!}{\includegraphics{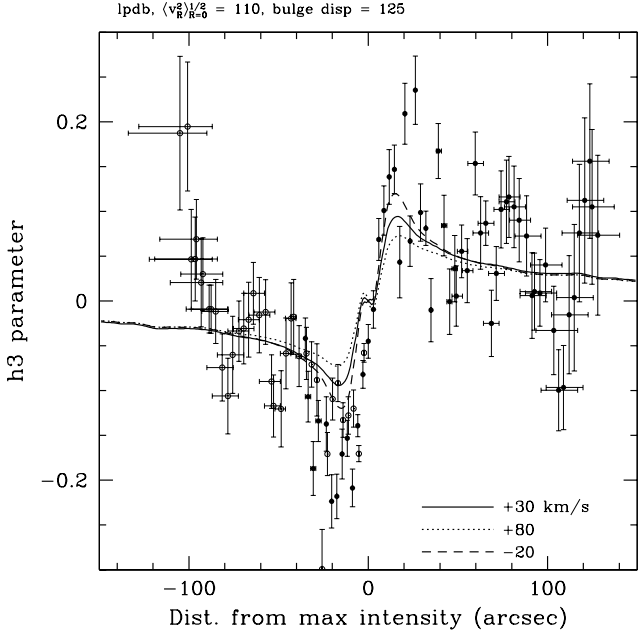}}
\caption[]{
As Fig. 12, but now for the asymmetry parameter.
A larger difference between bulge and disc rotation gives a larger
asymmetry of the profile. As a compromise between the fits
in figures 12 and 13 a bulge rotation of +30 \kmss has been
adopted, a value with an appreciable error. Nevertheless, in any case
the bulge needs a considerably lower rotation than the disc
to explain the observed asymmetric profiles.
}
\end{figure}

When the fit to the dispersion in Fig. 10 is considered in detail one can
note the following. For $R$ $\ga$ 25\arcsec\, by minimizing the difference
between model curves and data there is an excellent fit to the data
for $\disprn$ = 140 $\pm$ 30 \kms. However for $R$ $\la$ 25\arcsec\ one would
prefer a smaller disc dispersion: $\disprn$ = 110 \kmss
or even as low as 80 \kms. Then, not only the fit to the dispersion
is better, but also, as can be seen in Fig. 11, the line profiles
are more asymmetric resembling more what is observed. But can this
be done, such a smaller dispersion in the inner region?
In principle the dispersion over the whole radial extent is fixed
by the disc surface brightness through Eqs (6), (9), and (10).
But this assumes that the vertical scale parameter $(z_0)$ is the same
at all radii. If $z_0$ can be made smaller for $R$ $\la$ 25\arcsec\ then
the dispersion can also be made smaller while remaining
consistent with the photometry. This is not even such a bad idea,
in fact it is something which might be expected. Inwards of $R$ = 25\arcsec\
one is entering the bulge region where gas is sparce and star formation
has nearly ceased. Consequently there will be no spiral arms and no
molecular clouds and the disc is prohibited from heating up
(Lacey \cite{lacey91}). The disc then remains colder and thinner
than at those radii where there is star formation or where star formation
has existed longer. Such a lower value of $z_0$ at or near the bulge
would obviously go undetected in observations of edge-on galaxies.
Certainly the kinematic data of NGC 7331 can be explained better by a
thin and cold disc within the bulge. This fact, however, is not an
overwhelming evidence for making thin and cold discs within
bulges into a general hypothesis.

\begin{figure}
\resizebox{\hsize}{!}{\includegraphics{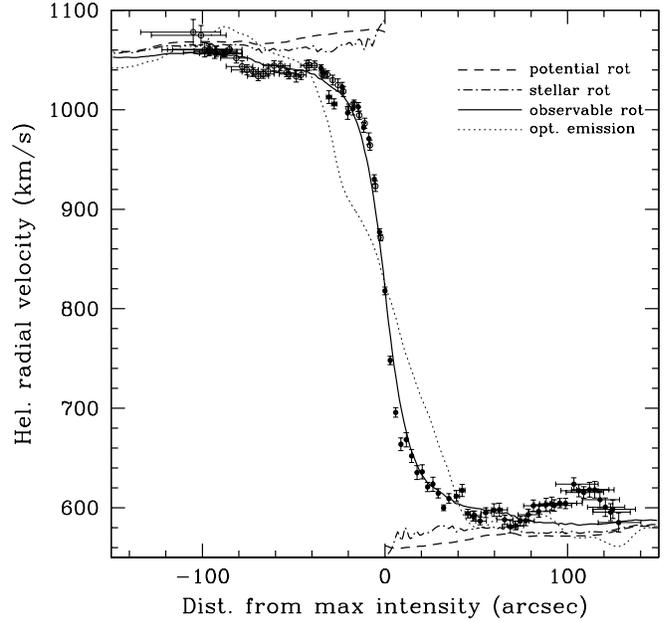}}
\caption[]{
Comparison of the radial velocities following from the model calculations
with the observed stellar radial velocities (for $\disprn$ = 140 \kms,
bulge rotation and dispersion of +30 and 125 \kmss respectively).
The input potential rotation was adapted to achieve the best fit.
}
\end{figure}

In Fig. 14 the radial velocities following from the model calculations
are compared with the observed radial velocities. As described above,
there is the freedom the choose the testparticle (potential)
input rotation such that the calculated radial velocities match
the observations. Thus a comparison of radial velocities cannot
give an extra constraint on the velocity dispersion of the
disc or on the bulge rotation. What can be concluded from
Fig. 14 is that a testparticle rotation curve can be constructed which
gives a consistent picture; in this case for $\disprn$ = 140 \kms.

The lpd bulge decomposition enables a reasonable
reconstruction of the observations. It is likely that an even better
fit can be made for slightly different bulge/disc
decompositions, for instance for bulges which
have brightness laws other than the $R^{1/4}$ employed so far.
But the parameter space which then opens up is so vast that
a huge amount of time is needed to investigate; while only giving
marginal improvements.

Some problems have to be discussed, however. Changing
the brightness ratio of model disc and bulge has a
considerable effect on the dispersions in the
disc dominated region. How well can one then constrain the
dispersion of the disc? Fortunately this appears to be a smaller
problem than expected. This is because the bulge/disc brightness
ratio cannot be changed at will; one has to remain compatible
with the observed photometry. In fact, during the construction of the lpd
bulge situation an iteration was made to increase the bulge
strength with a factor two at the transition radius. Bulge parameters and
disc surface brightness then had to be adapted to match the photometry.
Surprisingly, the derived disc dispersion turned out to be nearly
equal before and after this change. The error generated by
the not precisely determined bulge/disc brightness ratio is well
comprised within the 30 \kmss error on the value of $\disprn$.
Another problem is that the large observed dispersions near
$R$ = 100\arcsec\ cannot be explained. Looking at the fit to the
data in Fig. 10 this problem looks worse than it is because of
the strong oversampling of the data at that radius, caused by
the adopted method of averaging along the slit. Taking into
account the horizontal error bars which show the amount of oversampling,
there are only
four independent measurements (two on either side of the galaxy) which have
a large dispersion. Still these four measurements cannot be explained
by a regular galactic disc situation and consequently the disc is not regular.
A hotter disc by a factor of two locally at that radius
seems unlikely because then a very local increased heating mechanism
must have been present. More likely is a locally irregular
stellar velocity field generating broader profiles over
the 15\arcsec\ spatial resolution at those positions. Evidence supporting
this is the discrepant low apparent rotation at $R$ = 100\arcsec\
on the North side only.

\section{The amount of mass}
\subsection{The mass of the disc}
The surface density of the disc as a function of radius
$({\sigma}_{\rm d}(R))$ is given by

\begin{equation}
{\sigma}_{\rm d}(R) = {\sigma}_0^{\rm d} f(R) = 
{{( 0.6\; \disprn )^2 }\over{\pi G z_0 }} \cdot f(R).
\end{equation}
When substituting $\disprn = 140 \pm 30$ \kmss for the derived disc
dispersion and 7\arcsec\ = 506 pc for the value of $z_0$ a central surface
density ${\sigma}_0^{\rm d}$ of 1027 $\pm$ 440 $M_{\odot}$ pc$^{-2}$
is found. Using Eq. (8) and ${\mu}_0{\vert}_{\rm face-on} = 18.67$
I-mag. arcsec$^{-2}$ the mass-to-light ratio of the disc is
1.6 $\pm$ 0.7 $M_{\odot}/L_{\odot}^I$.

Some parameters entering into the calculation of surface density and
M/L ratio have not been measured but are inferred from other
studies. For example the factor 0.6 which might have a typical
error of 0.1; the value of $z_0$ having some 20\% error and there
is the uncertainty of the correction to face-on surface brightness
which is at present a simple cos(inclination) factor.
Nevertheless, NGC 7331 is a regular galaxy so that one may assume
that these parameters have been given the appropriate value.
The dominant source of errors is that of the dispersion anyway.

For a number of galactic discs stellar velocity dispersions have now 
been determined (Bottema \cite{bottema93}). Combining these results leads to
a disc M/L ratio of 1.8 $\pm$ 0.4 in the B-band (Bottema \cite{bottema97a}).
The value for NGC 7331 of 1.6 $\pm$ 0.7 in the I-band is then
consistent with the results for other galaxies under the likely
assumption that the B-I colour of the disc of NGC 7331 is
comparable to the solar value. This gives confidence that determined
and assumed parameters for NGC 7331 are correct.

From the surface density as a function of radius the rotation curve
has been calculated (Casertano \cite{casertano83}) which is shown in Fig. 15.
The maximum rotation contributed by the disc is 102 \kmss at
a radius of 3.8 kpc and the total mass of the stellar
disc amounts to 17.8 10$^9\; M_{\odot}$.

\begin{figure}
\resizebox{\hsize}{!}{\includegraphics{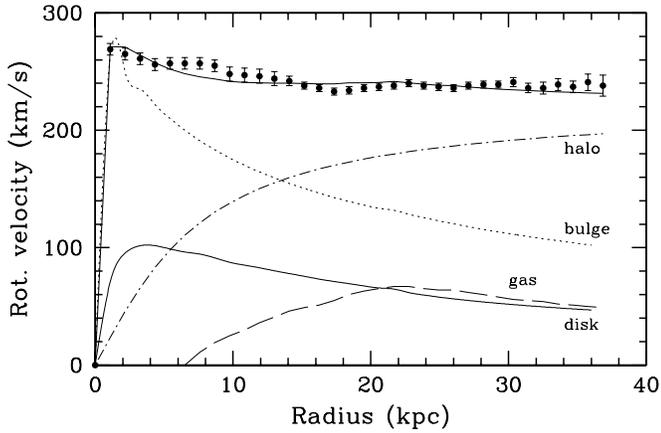}}
\caption[]{
Fit of the rotation curve of bulge and halo to the observed rotation
curve given by the dots. The rotation curve of the disc is fixed
by the observed stellar velocity dispersions.
Fit parameters are given in Table 4. As can be seen, the
bulge and halo dominate the mass distribution at all radii.
}
\end{figure}

\subsection{The rotation curve fit}
The rotation curve for NGC 7331 is presented by the dots in Fig. 15,
and is given by the \hi rotation for $R$ $\ga$ 70\arcsec\ and by the
iteration procedure to fit the stellar rotation for $R$ $\la$ 70\arcsec.
The contribution to the total rotation of the disc is fixed by the
observed dispersions, and the contribution of the gas is given by the
amount of \hi and He gas (Begeman et al. \cite{begeman91}). 
In order to obtain
the observed rotation a contribution of the bulge has to be added,
and to maintain the flat level at large radii a dark halo has to be
invoked. For this dark halo a spherical pseudo isothermal sphere
is assumed (Carignan \& Freeman \cite{carignan85}) with density law

\begin{equation}
\rho (r) = {\rho}_0^{\rm h} \left[ 1 + \left( {{r}\over{r_{\rm core}}}
\right)^2 \right]^{-1},\end{equation}
and rotation

\begin{equation}
v_{\rm h} (r) = v^{\rm h}_{\rm max} \sqrt{
1 - {{r_{\rm core}}\over{r}} {\rm arctan}\left(
{{r}\over{r_{\rm core}}} \right) },\end{equation}
where the maximum halo rotation $v^{\rm h}_{\rm max}$ and the
halo core radius $r_{\rm core}$ are two free parameters which can be
adjusted to match the observed rotation. A third free parameter is the
total mass of the bulge.

The surface brightness of the (lpd) bulge is given by Eq. (2) and
parameters of Table 2. This description is all right for $R \geq$ 2\arcsec\,
but, because of the steepness of the $R^{1/4}$ law near $R = 0$ care
was taken to use the observed surface brightness for $R <$ 2\arcsec.
For an assumed spherical density distribution a rotation curve for
the bulge can then be calculated from the surface brightness and
M/L ratio using the equation on page 1310 of Kent (\cite{kent86}). 
If the M/L ratio of the
bulge is taken equal to that of the disk (1.6 in I) then the maximum
rotation contributed by the bulge is 125 \kmss at $R \approx 1$ kpc.

A least squares procedure was used to fit the quadratic sum
of the rotation curves of disc, gas, bulge, and dark halo to the
observed rotation. The total mass of the bulge became fixed at
87.3 10$^9\; M_{\odot}$ if the bulge would extend until 
$R$ = 300\arcsec\ and at 77.7 10$^9\; M_{\odot}$ if it extends
until $R$ = 150\arcsec. For the latter, using the total light from
Table 3 one finds a M/L ratio of the bulge of 6.8 $M_{\odot}/L_{\odot}^I$,
being a factor 4.3 larger than the M/L ratio of the disc.
The free parameters of the halo became fixed at $r_{\rm core} =$
5.9 kpc and $v^{\rm h}_{\rm max} =$ 224 \kms. The fit of the sum of 
the rotation curves is shown in Fig. 15, together with the contributions
of the individual components.

Looking at Fig. 15 it is obvious that bulge and dark halo
dominate the potential for the whole galaxy and the disc only
provides a minor contribution. If the M/L ratios of bulge and
disc are fixed to be equal then, in addition to providing the rotation
at large radii, the halo also has to provide most of the rotation
at small radii, leading to a very small core radius. Such
a scenario is not very realistic. For the free fit of the three
parameters, the bulge mass is always determined nearly
completely by the innermost rotation points. Changing, for example,
the M/L ratio of the disc to 2.5, the maximum allowed by the error
on the dispersion, the total bulge mass only decreases from
77.7 to 72.9 10${}^9\; M_{\odot}$ but $(M/L)_{\rm b}/(M/L)_{\rm d}$
goes down to 2.6 instead of 4.3.
A summary of the relevant parameters and fitted values is
given in Table 4. There is at least one obvious conclusion
following from the rotation curve analysis and that is that
a bulge considerably more massive than the disc is needed,
having a M/L ratio which is also considerably larger than
the disc M/L.

\begin{table*}
\caption[]{Rotation curve fit parameters}
\begin{flushleft}
\begin{tabular}{llll}
\noalign{\smallskip}
\hline
\noalign{\smallskip}
 & lpd bulge & lpd bulge & small bulge \\
 & (M/L)$_{\rm d}$ = 1.58 & (M/L)$_{\rm d}$ = 2.5 & (M/L)$_{\rm d}$ =
1.93$^{(1)}$ \\
\noalign{\smallskip}\hline \noalign{\smallskip}
$M_{\rm disk}$ (10$^9$ $M_{\odot}$) & 17.8 & 28.1 & 25.0 \\
$v^{\rm disc}_{\rm max}$ (\kms) & 102 & 128 & 119 \\
$M_{\rm bulge}$ (10$^9$ $M_{\odot})^{(2)}$ & 77.7 & 72.9 & 40.5 \\
(M/L)$_{\rm b}$ & 6.8 & 6.4 & 5.0 \\
(M/L)$_{\rm b}$/(M/L)$_{\rm d}$ & 4.3 & 2.6 & 2.6 \\
$r_{\rm core}$ (kpc) & 5.9 & 7.5 & 3.0 \\
$v^{\rm h}_{\rm max}$ (\kms) & 224 & 232 & 219 \\
\noalign{\smallskip}
\hline
\noalign{\smallskip}
\multicolumn{4}{l}{\quad (1): All $M/L$ in $M_{\odot}/L_{\odot}^I$} \\
\multicolumn{4}{l}{\quad (2): until $R$ = 150\arcsec} \\
\end{tabular}
\end{flushleft}
\end{table*}

\subsection{The mass-to-light ratio of the bulge, a discussion}
For the lpd bulge situation a bulge mass-to-light ratio
of 6.8 is derived rather independent of the
mass of the disc. To obtain a reasonable error on this
value, the whole fitting procedure has been repeated for
the small bulge situation. This bulge is more compact and
hence less mass is needed to bring the rotation to the
observed values. In Table 4 the fitted values are presented;
the total bulge mass is 40.5 10$^9$ $M_{\odot}$ with a M/L ratio of
5.0. A small bulge situation is barely in agreement
with the observed dispersions and one may thus consider this M/L ratio of 5.0
as a two sigma deviation from the result for the
lpd bulge, leading to $(M/L)_{\rm b} = 6.8 \pm 1\;
M_{\odot}/L_{\odot}^I$.

How does this $(M/L)_{\rm b}$ ratio compare to other M/L ratios?
As noted above, it is larger by a factor around four
than the disc value of NGC 7331 and the disc value of
other galaxies for which stellar velocity dispersions have been
measured. But how about (M/L) ratios of bulges of other galaxies?
It appears that the literature gives essentially two methods to
determine a bulge M/L ratio. At first an application of
the Jeans equations to the observed velocity dispersions and morphology.
Secondly, when an emission line rotation curve is observed, a fit
to that curve can be made by assuming the bulge is the dominant
contributor to the rotation in the inner regions.

The first method has been used by Jarvis \& Freeman (\cite{jarvis85}) who
used Wilson (\cite{wilson75}) two integral distribution function models
to fit simultaneously the dispersion, rotation and photometry of
three galaxies with large and dominant bulges. This method
was also used by Kent (\cite{kent89}) according to the procedure
of Simien et al. (\cite{simien79}) 
which assumes stellar isotropy, to determine
the M/L ratio of the bulge of M31. 
For the second method, the rotation curve fit, one is referred to
the papers by Kent (\cite{kent86}, \cite{kent87}), 
from which I selected five
galaxies with substantial bulges and a well defined bulge/disc decomposition.
For the total of nine galactic bulges the M/L ratios are
given in Table 5. All listed M/L ratios are
converted to the I-band. For the first 3 galaxies
an average bulge V-I colour, $\langle V-I \rangle$ = 1.31
was assumed according to the colours of a
number of dominant bulges in the study of 
Peletier \& Balcells (\cite{peletier97}).
For the other bulges where a conversion from the r-band to I-band
is needed the colour for the M31 bulge was adopted, (r-I = 1.03)
to be a representative value. In summary, the conversions are
$(M/L)_I = 0.58 (M/L)_V = 0.42 (M/L)_B$ for the first 3 and
$(M/L)_I = 0.76 (M/L)_r$ for the others.
Solar absolute magnitudes are from Worthey (\cite{worthey94}) and all
mass-to-light ratios are converted to a Hubble constant of 75 \kmss 
Mpc$^{-1}$. 
The average $(M/L)_I$ value for the nine bulges in Table 5
is 5.5 $\pm$ 0.6, while the rotation curve fit for NGC 7331
gives $(M/L)_I$ = 6.8 $\pm$ 1. Although slightly large, the value
of NGC 7331 is in perfect agreement with the M/L's of other
bulges.

\begin{table}
\caption[]{Bulge M/L ratios}
\begin{flushleft}
\begin{tabular}{ll} 
\noalign{\smallskip}
\hline
\noalign{\smallskip}
Galaxy & (M/L)$_I$ \\
\noalign{\smallskip}\hline \noalign{\smallskip}
NGC 7814 & 6.1 \\
NGC 4594 & 4.4 \\
NGC 7123 & 4.6 \\
M31      & 3.8 \\
NGC 1353 & 4.8 \\
NGC 2608 & 3.5 \\
NGC 2815 & 7.8 \\
NGC 3200 & 6.0 \\
UGC 2885 & 8.2 \\
\noalign{\smallskip}
\hline
\end{tabular}
\end{flushleft}
\end{table}

Now that it has been determined that the
mass-to-light ratios of both the bulge and disc are in agreement
with M/L ratios of bulges and discs of other galaxies,
there is the fact that, in general, M/L's of bulges are larger
than those of discs. In numbers one has $(M/L)^I_{\rm disc} \approx
1.8 \pm 0.4$ and $(M/L)^I_{\rm bulge} = 5.5 \pm 0.6$, while in 
the B-band, $(M/L)^B_{\rm disc} = 1.8 \pm 0.4$ and $(M/L)^B_{\rm bulge}
\approx 13 \pm 1.5$, the difference is
even larger. Intuitively one would like to explain the difference
by a fading stellar population of the bulge. That is to say
when the bulge is older than the disc its M/L ratio should be
larger because the average stellar population is less luminous.
For instance Worthey (\cite{worthey94}) 
gives $(M/L)_I$ values for stellar populations.
Using solar abundances a population getting older
from age = 8 Gyrs to age = 17 Gyrs the $(M/L)_I$ value
increases by a factor 1.74. This may not be completely valid
for a bulge-disc comparison for various and obvious reasons.
Nevertheless this indicative factor 1.74 falls short by nearly
a factor of two to explain the observed difference in
M/L between bulges and discs. It would be interesting
to do a more appropriate population synthesis
to see if the observed factor can
be explained. If this proves to be impossible, a solution might be
to add dark matter, baryonic or non-baryonic, to the bulge.
This suggestion has already been put forward to explain the observed
large number of lensing events towards the Galactic bulge
(Alcock et al. \cite{alcock97}).

\section{General discussion and conclusions}
In B93, Fig. 1 disc velocity dispersions of 12 galaxies have been plotted
versus the absolute luminosity in B of the old disc population.
To include NGC 7331 in this sample, first its
old disc only absolute magnitude has to be calculated and
secondly the radial dispersion at one scalelength. The
absolute magnitude for the whole galaxy amounts to -21.73,
which is corrected for internal and Galactic extinction and taken from
Sandage \& Tammann (\cite{sandage81}), converted to a distance of 14.9 Mpc.
To subtract the light of the bulge the bulge/disc total light
ratio of 0.9 in the I-band was converted to a value of 0.56
in the B-band. For this an average bulge B-I colour of 2.3 mag. was
assumed, which is the average value of a number of large bulges
as observed by Peletier \& Balcells (\cite{peletier97}), and an average disc
B-I colour of 1.8, taken from de Jong \& van der Kruit (\cite{jong94}). 
To get the
absolute magnitude of the old disc only, 0.48 mag. was added
to the total galaxy value to subtract the bulge light. Then another
0.32 mag. was added to subtract the young disc light 
(Bottema \cite{bottema97a}) for
an estimated B-V disc colour of 0.8.
Finally we have $M^B_{\rm od} = -20.9$ for NGC 7331.

The radial velocity dispersion at any radius is given by
$\dispr = \disprn \cdot \sqrt{f(R)}$. To obtain the value at $R=h$, first
$h$ was determined as the average of $h$ = 43\arcsec\ (large bulge)
and $h$ = 62\arcsec\ (scalelength in outer regions) giving $h$ = 52\arcsec\
$\pm$ 10\arcsec. For $\disprn = 140 \pm 30$ \kmss one gets
$\disprh = 38 \pm 11$ \kms. If this value is plotted in Fig. 1 of B93
it turns out that the dispersion is low for its brightness;
nearly a factor of two lower than the dispersions
of other galactic discs. Now the errors are substantial and statistically
there is not a disagreement. Still one may wonder
about the cause of this low value; several possibilities exist.

Looking at Fig. 10, where the dispersions are given, at $R=h$ = 52\arcsec\
the observed dispersion is 45 $\pm$ 10 \kms. Assuming the
observed dispersion is approximately the tangential dispersion and
$\dispt / \dispr = {{1}\over{2}}\sqrt{2}$ (for a flat rotation curve),
the radial dispersion at $R=h$ should amount to 64 \kmss while in
reality the ``input'' or internal disc
value is 38 $\pm$ 11 \kms. How is that possible? A detailed consideration
of the modelling procedure shows that at first some 8 to 10 \kmss
can be added to the input value of 38 \kmss caused by integration effects.
The remaining 16 to 18 \kmss difference can be accounted for by
the bulge. This shows that the bulge still has a considerable
effect on the observable dispersion even at one radial scalelength,
and that $\disprh = 38$ \kmss is consistent with the observations.

An other reason for the low dispersion might be that the disc of NGC 7331
is not exponential and that instead of $h$ = 52\arcsec\ the scalelength that
should have been assumed is smaller, leading
to a larger value of $\disprh$. This demonstrates
the limitation of the parameterization of the data in Fig. 1 of B93.
Therefore presenting an M/L value is better and for NGC 7331
indeed, the derived $(M/L)_I$ of 1.6 is in agreement with values of
other galactic discs.
On the other hand, it cannot be excluded that a substantial
scatter in the $\disprh$/abs. mag. relation is intrinsic.
This might be caused by a range of edge-on aspect
ratios for equal mass discs as a  result of different
heating rates. That again may be the
result of different star formation histories or environment effects.
For example the massive bulge of NGC 7331 can have had
a profound stabilizing effect on the disc so that less
disc heating has appeared.

From the discussion above it is clear that in order
to investigate or measure the stellar velocity dispersion of discs
it is advisable to observe only galaxies
with small bulges. The sample of 12 discs is still small and
an extension is needed to make a more accurate determination of
the M/L ratio of discs. In addition, or maybe first because it is
relatively easy to do, it would be very useful to determine the aspect 
ratio (or $h/z_0$) of edge-on galaxies and to find
the functionality with disc mass and scatter onto that functionality.
That will give insight into the evolution and star formation
history of discs and allow a better interpretation of the disc
dispersion/luminosity relation.

Another matter to be discussed is the larger M/L ratio of bulges
compared to that of discs. If that is indeed true for most
galaxies, there are cosmological consequences. Because bulges
are relatively old and have undergone little recent star formation,
there must have been a considerable fading over the last 5 Gyrs.
That implies that around 5 Gyrs ago, typically at $z \sim 1$
the bulge to disc light ratio must have been larger than presently.
Such an effect should be detectable in deep field observations and
may compromise galaxy type classification for large look back times.

\medskip

Finally a compilation of the main conclusions:
\begin{enumerate}
\item
The determined kinematics of the ionized gas and of the stars
is regular and symmetric with respect to the centre of the galaxy.
\item
For $R <$ 40\arcsec\ the emission line gas appears to rotate slower than
the stars. A likely explanation is an inclined and warped
(w.r.t the plane) gas layer in the inner regions.
\item
In the bulge dominated region the absorption line profiles are asymmetric;
they have a shallow extension towards the systemic velocity.
\item
No counterrotating component is observed.
\item
Previous claims of a counterrotating component might be based on
a wrong interpretation of the data.
\item
A galaxy consisting of a bulge and disc with different kinematics
is able to explain the asymmetric stellar line profiles.
\item
A satisfactory fit to the observed kinematics is found
for a rapidly rotating disc with radially decreasing
dispersion and a slowly rotating bulge with a constant, 
125 \kmss dispersion.
\item
An even better fit can be made when the disc is relatively
thinner and colder in the bulge region compared to the disc
outside the bulge.
\item
The bulge influences the observable dispersion to well
beyond its light dominated region.
\item
The observed disc dispersion gives $(M/L)^I_{\rm disc} =
1.6 \pm 0.7$, consistent with previous determinations.
\item
The bulge mass-to-light ratio is $6.8 \pm 1\; M_{\odot}/L_{\odot}^I$.
\item
The disc contributes only in a minor way to the mass content
of NGC 7331.
\item
For a sample of discs and bulges one has on average
$(M/L)^I_{\rm bulge}/(M/L)^I_{\rm disc} = 3.0$ and converted to
the B-band $(M/L)^B_{\rm bulge}/(M/L)^B_{\rm disc} = 7.2$.
\end{enumerate}

\appendix
\section{An investigation of UGD and CCC}
\subsection{A short description of UGD}
Suppose the observed absorption line spectrum of a galaxy
is given by $G(\lambda )$ and the spectrum of a template
star representative of the galaxy's spectral content is
given by $S(\lambda )$.
Then $G({\lambda})$ can be described as a convolution of
$S({\lambda})$ with the line of sight velocity distribution,
or simply line profile $F(v)$:

\begin{equation}
G(\log \lambda ) = \int F(v) S(\log \lambda -v/c) dv,
\end{equation}
where $c$ is the speed of light and conversion to a $\log \lambda$ 
scale is needed to put
the spectrum on a linear velocity scale.
To retrieve the line profile $F(v)$ Eq. (A1) has to be inverted.
The UGD procedure is one of the methods to accomplish this. It assumes
that the line profile can be described by the sum of a collection of 
equidistant gaussian functions with a fixed dispersion. Then one can
solve for the amplitudes of the gaussians that yield a line profile
whose convolution with the stellar template $S$ gives the best fit
to the observed galaxy spectrum $G$. This solving is done by a least
squares fitting procedure which also gives the errors matching the
best fitting line profile. Suppose the dispersion of the gaussians is
$\Delta v$, then the separation of the gaussians should be not more
than $2\Delta v$. Obviously any features on scales less than $\Delta v$
cannot be retrieved. Here one also has the usual trade-off between
resolution and noise; a higher resolution gives a noisier profile and vice
versa.

\subsection{A short description of CCC}
The cross-correlation clean method (CCC) is described
in Sect 2.3. In order to invert Eq. (A1) an intermediate
step is used by calculating the cross-correlation function (ccf)
of the galaxy and template spectrum:

\begin{equation}
ccf = \int G(\log \lambda ) S(\log \lambda - {{v}\over{c}})
d\log \lambda . \end{equation}
It can be shown (Bottema et al. \cite{bottema87}) that the ccf is equal
to the convolution of the line profile and the auto-correlation
function of the template spectrum $acf_S$:

\begin{equation}
ccf = \int F(v) acf_S (\log \lambda - {{v}\over{c}}) dv.
\end{equation}
This has the advantage that all the absorption lines in the
spectrum have been combined in a single peaked ccf;
hence the noise is already considerably lower before any
fitting or deconvolution procedures
are applied. To retrieve the line profile $F(v)$ the ccf
is deconvolved using the clean method (H\"ogbom \cite{hogbom74}) with
the template $acf_S$ as beam (or point spread function). 
The resolution of the retrieved line profile is uniform
and is set explicitly. It can be as high as allowed
by the sampling criterion. Again one has the usual trade-off
between noise and resolution.

\subsection{Testing UGD}
When applied to spectra of the inner region of NGC 7331,
UGD produces a counterrotating component while CCC produces
only an asymmetric line profile. To investigate this discrepancy
both procedures have been tested on a few artificial
asymmetric line profiles. This testing scheme
is not meant to be an exhaustive investigation, but only to
highlight some aspects.

The average template spectrum used for the analysis of the NGC 7331
data (30.2 \kmss pixel$^{-1}$, 2040 pixels around 5100 \AA) was convolved
with an asymmetric line profile. Noise
was, or was not, added and the resulting spectrum was considered
to be the (test) galaxy spectrum. UGD was then applied
using the unconvolved spectrum as template and it was attempted
to retrieve the input line profile. The parameters of UGD were equal
to those used by P96 and used for the UGD analysis of NGC 7331
described in Sect 3.4; namely a gaussian dispersion of two
velocity pixels and a separation of 3 pixels.

\begin{figure}
\resizebox{\hsize}{!}{\includegraphics{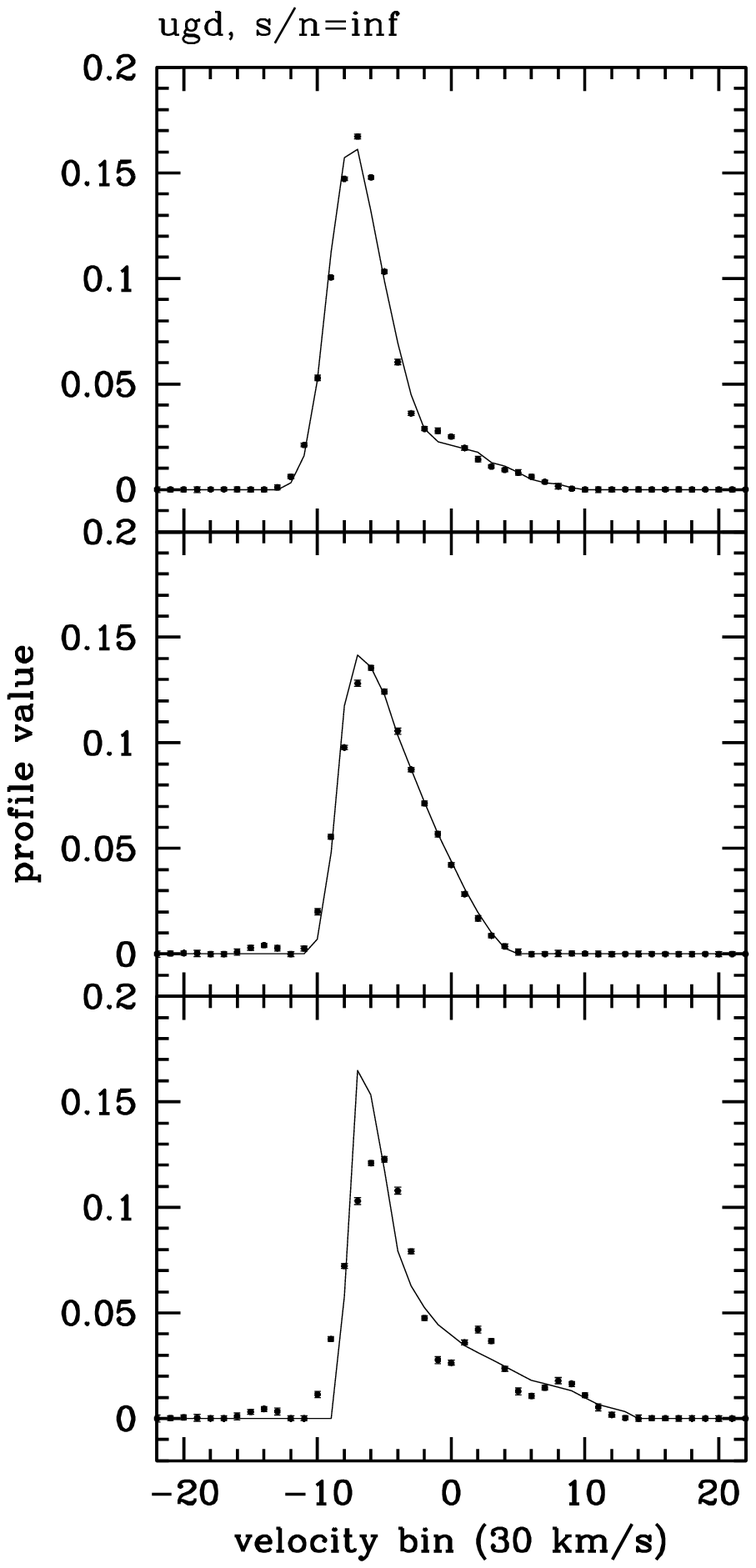}}
\caption[]{
Results of the line profile reconstruction using UGD for
three different asymmetric input profiles given by the lines.
The UGD result is given by the dots. In this case
no noise was added to the galaxy spectrum. When
the profile is broader than the preset resolution of the method
a good reconstruction is obtained. Otherwise (bottom panel)
aliases appear.
}
\end{figure}

For galaxy spectra with no noise added results for three different
line profiles are shown in Fig. A1; where the input profile as given by
the line is compared with the profile determined by UGD. Formal
fitting errors are superposed on the fit, but these are
small in this case. In the top panel the input line profile
is equal to that generated by the line profile calculation procedure
described in Sect. 5, for NGC 7331 with an lpd bulge, $\disprn$ =
110 \kms, $v^{\rm rot}_{\rm bulge} = +30$ \kms, at a distance of 20\arcsec\
from the nucleus. At that radial distance the observed counterrotating
component stands out very clearly. 
In the middle panel of Fig. A1 a relatively broad profile
is shown, while the lower panel depicts the fitting result
to a narrow profile with a long one sided extension.
As can be seen, UGD gives a good reconstruction of the profiles
in the upper and middle panels. The profile in the lower panel, however,
is badly reproduced; next to the main profile there
have appeared two distinct aliasing features not present in the input profile.
But here some care should have been taken that has not been.
A part of the input profile is narrower than the UGD resolution and hence
UGD can never reproduce the profile. Taking a set of narrower
gaussians improves the reconstruction. Unfortunately
in practice one has the problem that it is not known in advance
how narrow any galactic profile is, consequently making a suitable
choice of UGD parameters is not straightforward.

Next noise was added to the artificial galaxy spectra for a S/N 
level of 20. Results of the UGD reconstruction procedure for the
same three input profiles are shown in Fig A2. Now
also for the profiles in the top and middle panel the fit starts to
deviate from the input; aliasing features are produced on the
wing side of the profile. These aliasing features are equal
to what may be interpreted as a counterrotating component in
a bulge-disc situation. Remember that the top profile is equal to
what actually may be expected to occur for NGC 7331
in the ``counterrotating'' region. Considering the errors, however,
these aliasing features are not significant. The correct interpretation
of the UGD result should be to disregard these features, but
to arrive at this correct interpretation the fit errors have to be
presented. When one has a number of subsequent skewed profiles
all having such aliases produced by UGD
and no erros are given, a counterrotating component seems
to be present, while in reality it is an artifact. This is what may
have misled P96 to postulate a counterrotating bulge in NGC 7331.

\begin{figure}
\resizebox{\hsize}{!}{\includegraphics{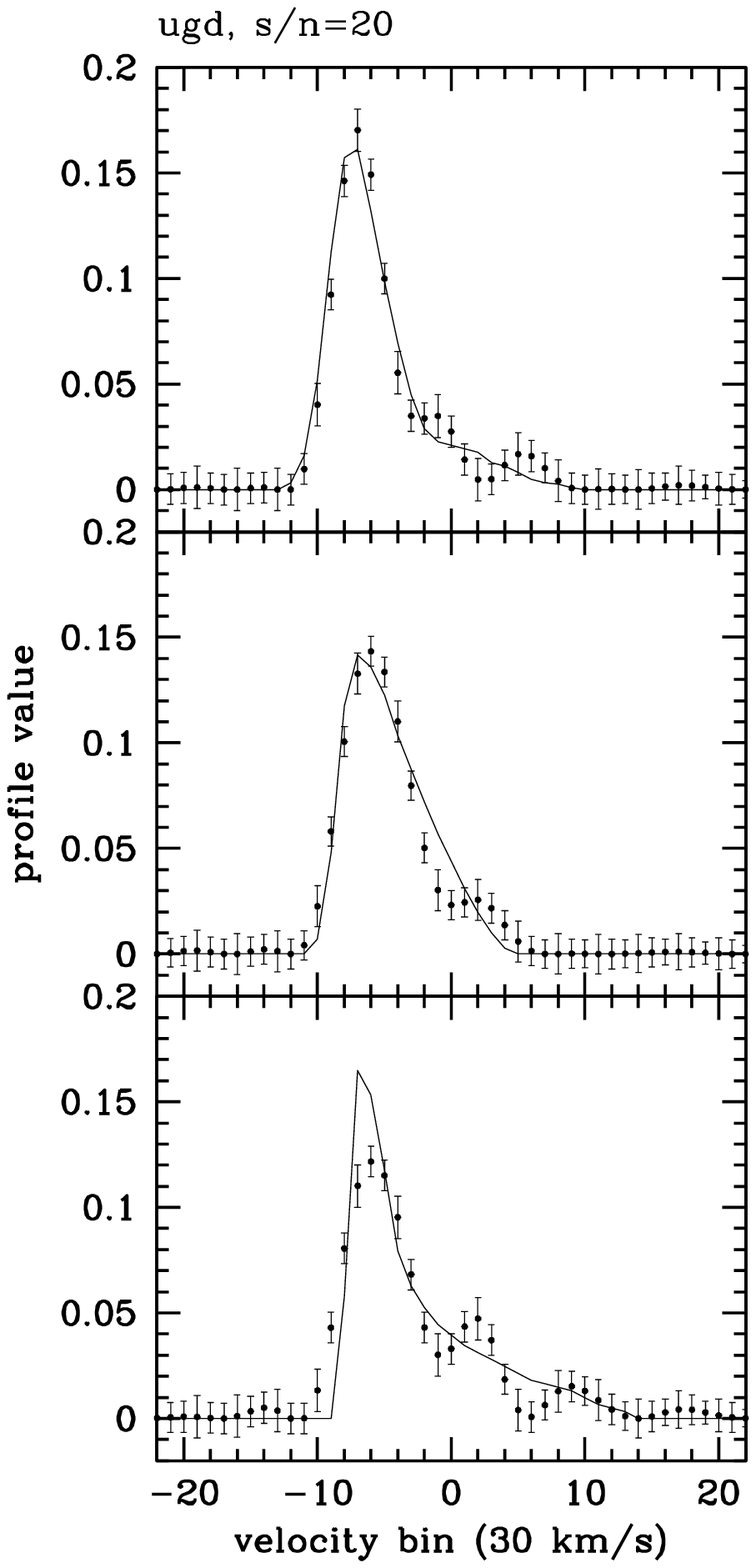}}
\caption[]{
As Fig. A1, but now for noise added to the galaxy spectrum
for a S/N level of 20. In all cases aliasing is present, though
considering the errors, not at a significant level. However,
when the errors are omitted one can be made to believe
that additional (counterrotating) line profile components are present.
}
\end{figure}

Lowering the S/N level in general increases the magnitude
of the aliases. This can be seen, for example, when comparing
the lower panels of figures A1 and A2 where the too narrow profile is
presented. A short investigation has been done to see if the UGD
fit may be improved when the width and separations of the
gaussians are decreased. To that aim $\Delta v$ was put at 
1${{1}\over{2}}$ pixel and the separation at 2 pixels and the fitting
procedure was rerun for the same profiles and S/N levels. For
the no-noise case the reconstruction of the profile in the bottom
panel has improved although still some significant aliasing remains. The
other two profiles, however, have become somewhat more irregular.
When noise is put on for a S/N = 20, in all cases substantial aliasing
appears accompanied with a substantial increase of the errors
of the fitted profiles. If one had to judge the reconstructed profiles,
one would have concluded that the S/N is too low, or a set of wider
gaussians is needed.

\subsection{The applicability of UGD}
Many more tests could have been performed. For instance for
different noise levels, for different dispersions and separations of the
gaussians, and the effects of template mismatch could have been
investigated using an additional different template spectrum.
To some this will be interesting but for the present case
an adequate explanation of the appearing features has been given.

In fact, the UGD testing as described above
is complementary to the test results that KM93
obtained in their original paper describing the method. From that paper
it is already clear that problems may arise for skewed profiles.
Therefore Kuijken and Merrifield issue two clear
warnings when applying UGD:
\medskip
\begin{enumerate}
\item
``The errors always have to be superposed on the fit
especially for noisy data such that a careful analysis can be made''
\item
``The size of the errorbars can be reduced by making the gaussian
components broader with a larger separation. However, when
the profile is narrower than the gaussians spurious features
may be introduced''
\end{enumerate}
In practice the second warning will always cause a problem
because one does not know in advance the dispersion
of the stellar population or subpopulations investigated.
So any small feature generated by UGD needs a further
investigation to establish its reality.
An additional problem associated with UGD is its hybrid
resolution. Features with dispersion less than $\Delta v$ are 
unresolved while broader features are resolved. This may cause
problems when interpreting the data.

\subsection{Testing CCC}
For the same template spectrum and the same three line profiles
the CCC method has been investigated. The ccf was
calculated between test galaxy and template. This ccf was
then cleaned, subtracting components until below the noise level.
The line profile was restored by convolving the components with
a gaussian having a resolution of 2${{1}\over{2}}$ velocity pixels
(as for the data analysis of NGC 7331) and by adding the residuals.
The errors were determined by measuring the standard deviation
in a region on the sides of the
profile. That value was assumed to be a representative error
for all pixels of the profile. The reconstructed profile should now be
equal to the input profile convolved with a gaussian of 2${{1}\over{2}}$
pixels. In Fig. A3 this comparison is made for a S/N level of 20.
As can be seen, the agreement between input and reconstructed profile
is very satisfactory for all three cases. The errors are
considerably smaller than those of UGD for the same S/N ratio
although a comparison with UGD is not completely fair because
UGD tries to reconstruct the profile at full resolution, CCC not.
In the middle panel of Fig. A3 one may notice that CCC
has the same tendency as UGD to make the profile
a bit too narrow in the top. On the other hand, the bottom
panel shows that CCC does not give any problems reconstructing
extremely narrow and skewed profiles.

\begin{figure}
\resizebox{\hsize}{!}{\includegraphics{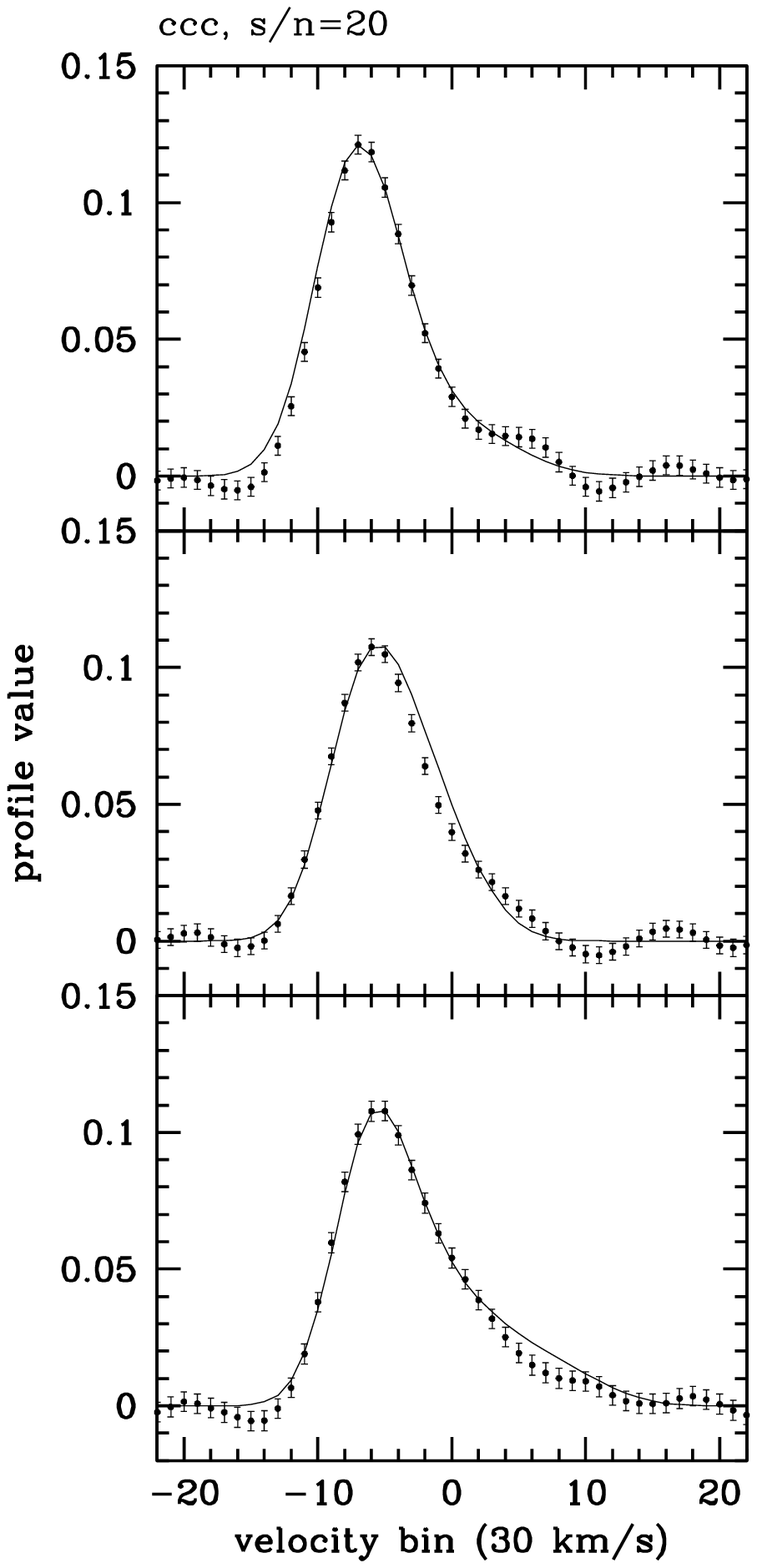}}
\caption[]{
Results of the line profile reconstruction using CCC
for the same three input profiles as in figures A1 and A2.
CCC can only retrieve the profiles convolved with a certain
preset resolution. The output profile is
thus compared with the input profile convolved to that resolution.
In all cases a good reconstruction is achieved.
}
\end{figure}

Naturally CCC has its disadvantages; the most obvious being
that the reconstructed profile is always convolved
to a certain (but uniform) resolution.
The biggest advantage of the method is that it can be applied
without any prior knowledge of the actual galactic line profile.
In addition it is straightforward in the sense that no
iterative fitting is involved.

\begin{acknowledgements}
The observations presented in this paper are obtained with the
Isaac Newton Telescope which is operated on the island of La Palma
by the Isaac Newton Group in the Spanish Observatorio
del Roque de los Muchachos of the Instituto de
Astrofisica de Canarias.
Many thanks to Jan Lub, who did the actual observations
during my unforeseen absence.
I also thank Konrad Kuijken for reading the manuscript
and I thank him and Joris Gerssen for helpful discussions.
The Kapteyn Institute is acknowledged
for hospitality and support. 
\end{acknowledgements}

\end{document}